\documentclass{article}
\PassOptionsToPackage{numbers,compress}{natbib}
\usepackage[main,final]{neurips_2026}
\usepackage{placeins}
\usepackage[utf8]{inputenc}
\usepackage[T1]{fontenc}
\ifdefined\XeTeXversion
  \catcode"2019=13
  \def’{'}
\fi

\usepackage{url}
\usepackage{graphicx}
\usepackage{xcolor}
\usepackage{amsmath,amsfonts,amssymb,amsthm}
\usepackage[ruled,vlined,linesnumbered]{algorithm2e}
\usepackage{xspace}
\usepackage{booktabs}

\newtheorem{assumption}{Assumption}
\newtheorem{theorem}{Theorem}

\newcommand{\sysname}{\textsf{Aegis}\xspace}

\title{Aegis: Generative Gradient Masking for Privacy-Preserving Medical Federated Learning}
\author{%
  \parbox[t]{\dimexpr\textwidth-2\leftmargini\relax}{%
    \centering
    \textbf{%
      Chaoyu~Zhang\textsuperscript{1},
      Shanghao~Shi\textsuperscript{2},
      Heng~Jin\textsuperscript{1},
      Ning~Wang\textsuperscript{3},\\
      Y.~Thomas~Hou\textsuperscript{1},
      Wenjing~Lou\textsuperscript{1}%
    }\endgraf
    \normalfont
    \mbox{\textsuperscript{1}Virginia Tech}\quad
    \mbox{\textsuperscript{2}Washington University in St. Louis}\quad
    \mbox{\textsuperscript{3}University of South Florida}\endgraf
  }%
}

\begin{document}
\maketitle
\begin{abstract}

Federated learning (FL) has become a foundational paradigm for multi-institutional medical AI, allowing hospitals and research centers to jointly train diagnostic models without exchanging patient records. This privacy promise, however, is increasingly contested: a malicious or honest-but-curious server can launch model inversion attacks (MIAs) that reconstruct private patient images directly from shared model updates, and recent scalable, closed-form attacks penetrate even secure aggregation at clinically realistic batch sizes. Existing defenses face an unsatisfactory dilemma. Gradient-perturbation methods such as differential privacy and pruning trade away the diagnostic accuracy on which clinical reliability depends, while cryptographic protocols add system complexity yet still leave updates exposed to these scalable attacks. We propose \sysname, a principled client-side defense that breaks this dilemma without perturbing patient data or modifying the FL protocol. Our key insight is that the success of every known MIA is fundamentally bounded by the local batch size relative to the model's leakage capacity; once this limit is exceeded, distinct samples collide and reconstructions collapse into indistinguishable mixtures. \sysname turns this universal bottleneck into a defense: each client superimposes onto its real update a masking gradient computed on locally synthesized, task-relevant data, deliberately pushing the effective batch beyond the attack's recovery capacity. We complement the design with theoretical convergence guarantees under standard convex assumptions and evaluate \sysname on MNIST, CIFAR-10, and three MedMNIST modalities (chest X-ray, abdominal CT, colon pathology). \sysname neutralizes three state-of-the-art MIAs while preserving model utility and incurring only modest overhead, offering a practical privacy primitive for medical FL.

\end{abstract}

\section{Introduction}\label{introduction}

FL has emerged as a foundational paradigm for multi-institutional medical collaboration: hospitals, research centers, and biotech labs can jointly train diagnostic and biomedical models without ever moving raw patient data, because only model updates (weights or gradients) are transmitted to a central server~\cite{mcmahan2017communication}. This decentralization is precisely what makes FL attractive for clinical AI, where data is fragmented across silos governed by HIPAA~\cite{moore2019review}, GDPR~\cite{li2019impact}, and institutional review boards. Yet the privacy guarantee of FL is increasingly contested. MIAs, launched by an honest-but-curious or actively malicious server, can reconstruct private client samples directly from the shared updates. A first generation of optimization-based MIAs~\cite{zhu2019deep,zhao2020idlg,geiping2020inverting,zhu2020r,yin2021see,lu2022april,hatamizadeh2022gradvit} iteratively matches dummy gradients to real ones, raising particularly serious concerns when the data being reconstructed are sensitive medical images such as CT slices, chest X-rays, or histopathology tiles.

To mitigate MIAs, existing defenses fall into two categories: gradient perturbation and cryptographic protection. Gradient-perturbation methods, exemplified by differential privacy (DP)~\cite{46432}, inject noise into local gradients to obscure sensitive signal at the cost of model accuracy, a cost that is particularly damaging in medical applications, where small accuracy drops can translate into clinically meaningful errors. Adaptive variants~\cite{wang2023more, wang2022protect} calibrate noise to estimated leakage risk, and pruning-based approaches~\cite{zhu2019deep, sun2021soteria} selectively suppress recoverable gradient components. Cryptographic defenses based on multi-party computation~\cite{knott2021crypten} implement secure aggregation protocols~\cite{bell2020secure, guo2020v, kadhe2020fastsecagg, xu2019verifynet} so that the server only ever sees aggregated updates.

However, a recent and rapidly growing class of MIAs~\cite{zhao2023loki, shi2023scale, wen2022fishing, pasquini2022eluding, fowl2021robbing, shi2025medleak} demonstrates strong reconstruction even \emph{under} secure aggregation. By exploiting the \emph{linear-leakage primitive}~\cite{fowl2021robbing}, these attacks bypass iterative optimization and recover inputs in closed form, scaling to large batches and remaining effective in production FL settings. Related leakage risks also arise in federated language-model fine-tuning~\cite{shi2026efficiency} and split learning~\cite{jin2026privacy}. Consequently, current defenses face an unsatisfactory dilemma: perturbation-based methods sacrifice the model utility on which clinical reliability depends, while cryptographic approaches add system overhead yet do not eliminate the leakage. This gap is critical for medical FL: if linear-leakage MIAs can reconstruct patient scans from shared updates, the privacy promise that justifies federated training in the first place is broken. We therefore set out to design a defense that (i) is effective against scalable, linear-leakage MIAs, (ii) preserves model utility on clinical data, and (iii) requires no modification to existing FL protocols.

\textbf{Our solution.} We introduce \sysname, a generative-data-based defense against MIAs. Our key observation (Tab.~\ref{tab:literature-review}; Appendix~\ref{appendix:mia_survey}) is that \emph{the reconstruction success of every known MIA is bounded by the local batch size relative to model capacity}. Optimization-based MIAs degrade rapidly beyond very small batches; linear-leakage attacks scale further but, by their own mathematical structure, succeed only when the local batch size does not exceed the number of neurons in the first fully connected layer~\cite{shi2023scale}. Once that capacity is exceeded, multiple inputs collide in the same leakage bins and reconstructions collapse into mixed, indistinguishable images.

\sysname turns this fundamental bottleneck into a defense. Each client uses an off-the-shelf generative model to synthesize an auxiliary \emph{defense dataset} large enough that, when combined with the real local batch, the effective batch size exceeds the leakage capacity of the targeted layer. After standard local training, the client computes an additional \emph{masking gradient} on the defense data and superimposes it onto its real update before sharing. Crucially, \sysname does not perturb real patient samples, drop informative gradients, or alter the FL protocol; it exploits the structural limit of linear leakage so that any closed-form reconstruction the server attempts collapses into a blended mixture. We trade a moderate, one-time generation cost and an extra masking pass per local epoch for strong empirical privacy at a small utility cost, yielding a practical privacy-preserving building block for medical FL.

\textbf{Contributions.} (i) We identify local batch size relative to first-FC-layer capacity as a unifying bottleneck for state-of-the-art MIAs, including those that defeat secure aggregation. (ii) Building on this observation, we design \sysname, a principled, protocol-compatible defense that masks real gradients with large-batch gradients computed on locally synthesized task-relevant data. (iii) We provide a convergence analysis under standard convex assumptions that quantifies how the discrepancy between real and synthetic data enters the convergence bound; we are explicit that this is not a non-convex deep-learning guarantee. (iv) On MNIST, CIFAR-10, and three MedMNIST modalities (chest X-ray, abdominal CT, colon pathology), \sysname suppresses Robbing-the-Fed, LOKI, and Scale-MIA while matching or exceeding the model utility of state-of-the-art defenses, with running time competitive with lightweight baselines.

\section{Preliminaries and Related Work}\label{LL}

\textbf{Federated learning.} An FL system trains a global model $G_\theta$ across a set of distributed clients $\mathcal{C} = \{c_i\}_{i=1}^N$, each holding a private dataset $D_i$ (e.g., a hospital's patient images). At each round $t$, a subset $\mathcal{C}^t \subseteq \mathcal{C}$ either uploads gradients (\emph{FedSGD}), $\theta^{t+1} = \theta^{t} - \eta \sum_{c_i \in \mathcal{C}^t} p_i\, \nabla \mathcal{L}\bigl(\theta^{t}, D_i\bigr)$,
or post-training local-parameter deltas $\delta_i^{t}$ (\emph{FedAvg}), $\theta^{t+1} = \theta^{t} + \sum_{c_i \in \mathcal{C}^t} p_i\, \delta_i^{t}$,
where $\eta$ is the global learning rate, $\{p_i\}$ are non-negative client weights summing to one over $\mathcal{C}^t$, and $\mathcal{L}$ is the loss. While only updates leave each client, recent work shows that these updates are themselves an attack surface: adversaries can reconstruct $D_i$ via MIAs.

\subsection{Model Inversion Attacks in Federated Learning}

MIAs in FL fall into two families.

\begin{figure}[t]
\centering
\includegraphics[width=0.75\linewidth]{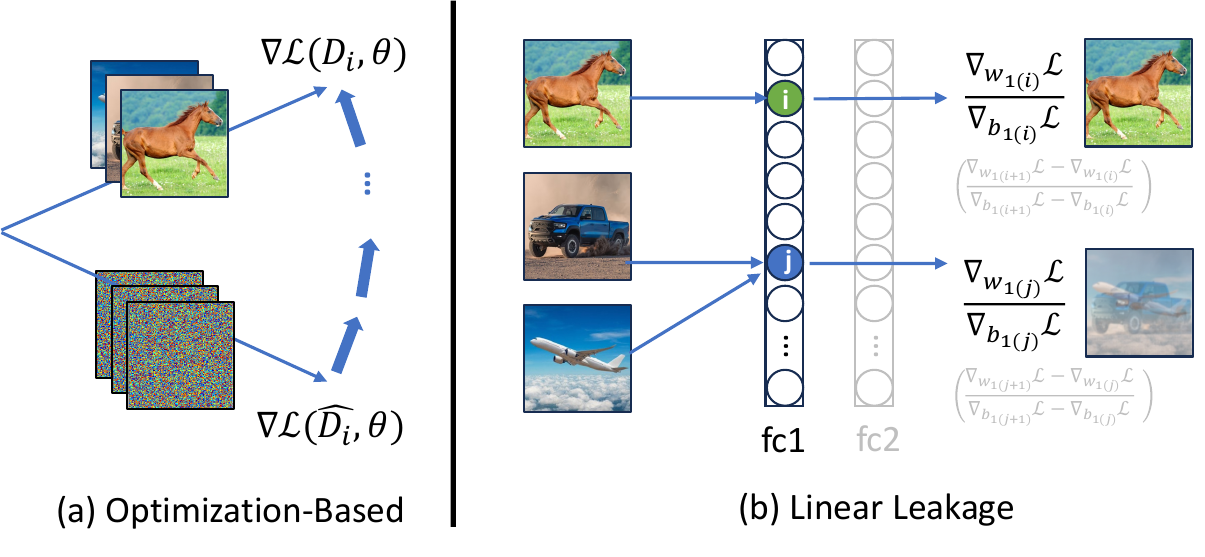}

\caption{Two families of MIAs in FL: (a) optimization-based attacks iteratively minimize the distance between dummy and target gradients; (b) linear-leakage-based attacks recover inputs in closed form from the weight and bias gradients of a fully connected layer.}

\label{Opt_LL}
\vspace{-0.5cm}
\end{figure}

\textbf{Optimization-based attacks} reconstruct local data by solving
$D_i^* = \arg\min_{\hat{D}_i} \lVert \nabla \mathcal{L}(\hat{D}_i, \theta) - \nabla \mathcal{L}(D_i, \theta) \rVert$
to match dummy gradients to the observed ones (Fig.~\ref{Opt_LL}(a)). DLG~\cite{zhu2019deep} pioneered this line, refined by iDLG~\cite{zhao2020idlg}, Inverting Gradients~\cite{geiping2020inverting} for higher fidelity, and GradInversion~\cite{yin2021see} for batched inputs on ResNet. In production-scale FL settings (large local batch sizes, multiple local epochs, realistic dataset sizes, and non-IID partitions), the success of these attacks degrades sharply, and a wide range of defenses can counter them~\cite{wang2023more}.

\textbf{Linear-leakage-based attacks} bypass iterative optimization and recover inputs in closed form, dramatically improving feasibility under realistic FL conditions and even under secure aggregation (Fig.~\ref{Opt_LL}(b)).

\textbf{The linear-leakage primitive~\cite{aono2017privacy}.} For a fully connected (FC) layer $y = w_1 x + b_1$ with $x \in \mathbb{R}^d$, $y \in \mathbb{R}^k$, $w_1 \in \mathbb{R}^{k \times d}$, $b_1 \in \mathbb{R}^k$, if a single input $x$ activates neuron $i$ in isolation, then it can be recovered exactly as $x = \nabla_{w_1(i)}\mathcal{L} / \nabla_{b_1(i)}\mathcal{L}$. When multiple inputs activate the same neuron, the corresponding gradient becomes a mixture, and the recovered image is a superposition of all activating inputs (Fig.~\ref{Opt_LL}(b), blue arrows).

Robbing-the-Fed~\cite{fowl2021robbing} turns this primitive into a scalable attack by treating each neuron of a crafted FC layer as a \emph{leakage bin}: weights $w_1$ encode a known cumulative distribution function over input statistics (e.g., image brightness), and biases $b_1$ act as thresholds, so that with negative biases and positive weights the ReLU activates exactly the bins whose threshold is crossed by a given input. Two consecutive FC layers $y = \mathrm{ReLU}(w_1 x + b_1)$ and $z = w_2 y + b_2$ enable the closed-form recovery $x_i = \bigl(\nabla_{w_1(i+1)}\mathcal{L} - \nabla_{w_1(i)}\mathcal{L}\bigr) / \bigl(\nabla_{b_1(i+1)}\mathcal{L} - \nabla_{b_1(i)}\mathcal{L}\bigr)$,
which holds whenever the second FC layer is configured so that $\nabla_{w_2(i)}\mathcal{L}$ and $\nabla_{b_2(i)}\mathcal{L}$ are equal across neurons activated by the same input. This primitive has since been extended to break secure aggregation (Fishing for Data~\cite{wen2022fishing}, LOKI~\cite{zhao2023loki}, Scale-MIA~\cite{shi2023scale}), to convolutional architectures (R-GAP~\cite{zhu2020r}), and to vision transformers (APRIL~\cite{lu2022april}).

\textbf{The structural bottleneck.} Critically, linear-leakage attacks succeed only while the local batch size $B$ does not exceed the number of leakage bins, i.e., the number of neurons $k$ in the first crafted FC layer~\cite{shi2023scale}. Once $B > k$, multiple inputs unavoidably activate the same bin, the gradient ratios above become mixtures, and reconstructions degrade into blurry, ghost-laden composites. \emph{We therefore identify the ratio of effective batch size to first-FC-layer capacity as the key factor governing the success of linear-leakage MIAs, and we take this bottleneck as the target of our defense.}

\subsection{Defenses against MIAs}

\textbf{Perturbation-based defenses.} Early defenses applied gradient compression~\cite{zhu2019deep} or DP~\cite{46432}; subsequent work~\cite{wang2023more, wang2022protect} argued that such generic mechanisms are not specifically targeted at MIAs and incur a noticeable privacy and utility trade-off. Soteria~\cite{sun2021soteria} perturbs FC-layer representations based on the observation that leakage originates in latent space; GradDefense~\cite{wang2022protect} extends perturbation to all layers using a sensitivity metric; Outpost~\cite{wang2023more} provides a lightweight, calibrated noise-addition scheme. All of these defenses, however, alter the very gradients used for learning and therefore degrade utility, an outcome that is especially problematic for clinical models.

\textbf{Cryptographic defenses.} Secure aggregation (SA) protocols~\cite{bonawitz2017practical, bell2020secure, guo2020v, kadhe2020fastsecagg, xu2019verifynet} built on multi-party computation~\cite{knott2021crypten} mask local updates with random bits while preserving the aggregate, $\mathrm{SA}(\delta_1, \dots, \delta_N) = (u_1, \dots, u_N)$ s.t. $\sum_{i=1}^{N} u_i = \sum_{i=1}^{N} \delta_i$. SA preserves utility but adds nontrivial system and communication overhead. More importantly, recent attacks defeat it directly: gradient-disaggregation~\cite{wen2022fishing} infers individual updates from aggregates with crafted models, gradient-suppression~\cite{pasquini2022eluding} zeros out all but a target client's update, and large-batch-reconstruction~\cite{fowl2021robbing} together with LOKI~\cite{zhao2023loki} (which extends linear leakage to convolutional modules under FedAvg) reconstruct global batches in closed form even after aggregation.

\textbf{Generation-based defenses.} Gong~et~al.~\cite{gong2023gan} explored generated-data-based defenses against GAN-based model inversion attacks in the centralized setting, biasing the adversary to recover synthesized rather than original samples. Their setup, however, is centralized rather than federated and targets a different attack surface; we are not aware of prior FL-specific defenses that exploit linear-leakage's structural bottleneck via auxiliary generated data.

\subsection{Generative Models for Image Synthesis}

A generative model trained or queried with class- or text-conditioning provides the auxiliary in-distribution data that \sysname requires. Modern conditional image generators include GANs~\cite{goodfellow2014generativeadversarialnetworks} (e.g., StackGAN~\cite{zhang2017stackgan}, AttnGAN~\cite{xu2018attngan}), VAEs~\cite{kingma2013auto}, and diffusion-based models~\cite{ho2020denoising,Rombach_2022_CVPR,saharia2022photorealistic,ramesh2022hierarchical,betker2023improving}, the latter often combined with text encoders such as CLIP~\cite{radford2021learning} or T5~\cite{raffel2020exploring}. \sysname is not tied to a specific generative architecture; it requires that the synthesized auxiliary samples are sufficiently task-relevant to avoid harming utility while still inducing gradient collisions in the leakage bins (see Sec.~\ref{sec:system-design}).

\section{Threat Model}\label{sec:threat-model}

We study a multi-institutional medical FL system where clients $\{c_i\}_{i=1}^{N}$, such as hospitals or research centers, jointly train a diagnostic model under a server. Each client keeps private data $D_i$ (e.g., CT slices, chest X-rays, or histopathology tiles) locally and sends only updates $\delta_i$. The server defines the task and protocol, coordinates training, and may act as the strongest known active MIA adversary~\cite{fowl2021robbing, zhao2023loki, shi2023scale}. Its goal is to reconstruct the private data union $\bigcup_{i=1}^{N} D_i$ from received updates.

\textbf{Adversary capabilities.} The server knows and controls the global model $G$ and parameters $\theta$ at every round. It may alter the architecture or initialization to plant leakage primitives, such as crafted FC layers calibrated for linear leakage~\cite{fowl2021robbing}, and may attack at any round, especially early rounds when random parameters hide malicious structure and the model has not converged. It can also approximate the global data distribution using public data, self-collected samples, or colluding clients, yielding an auxiliary set $D_{\mathrm{Adv}}$ for bin calibration. The adversary knows standard FL primitives, including FedSGD, FedAvg, and secure aggregation, and mounts linear leakage on top of them as in current state-of-the-art MIAs~\cite{fowl2021robbing, zhao2023loki, shi2023scale}.

\textbf{Defender model.} The defenders are the legitimate medical clients. They aim to protect patient data while sending useful updates, so diagnostic accuracy should stay close to the no-defense baseline. Each client sees only the broadcast model and cannot reliably tell whether it is benign or crafted as $G_{\mathrm{adv}}$, nor which MIA variant is being used. Existing defenses such as SA, DP, and gradient pruning either remain vulnerable to scalable linear leakage or incur utility losses that are problematic for clinical use (Sec.~\ref{EE}).

\textbf{Why local hyperparameters alone do not suffice.} One might try to stop MIAs by changing local training. However, with Robbing-the-Fed~\cite{fowl2021robbing} as a representative attack, varying batch size, learning rate, and local epochs within utility-preserving ranges does not meaningfully reduce reconstruction quality (Tab.~\ref{tab:local}; Appendix~\ref{appendix:local_robbing}). This motivates an explicit \emph{defense}, not reliance on training hyperparameters.

\section{System Design}\label{sec:system-design}

\textit{Threat-model recap (Sec.~\ref{sec:threat-model}).} An active server may craft the broadcast model $G_{\mathrm{adv}}$ to enable closed-form linear-leakage reconstruction of client images; the client controls only its local training and what it transmits to the server, and cannot reliably distinguish a benign global model from a maliciously crafted one. \sysname is a client-side mechanism that operates entirely within this constraint.

\begin{figure*}[t]
\centering
\includegraphics[width=\linewidth]{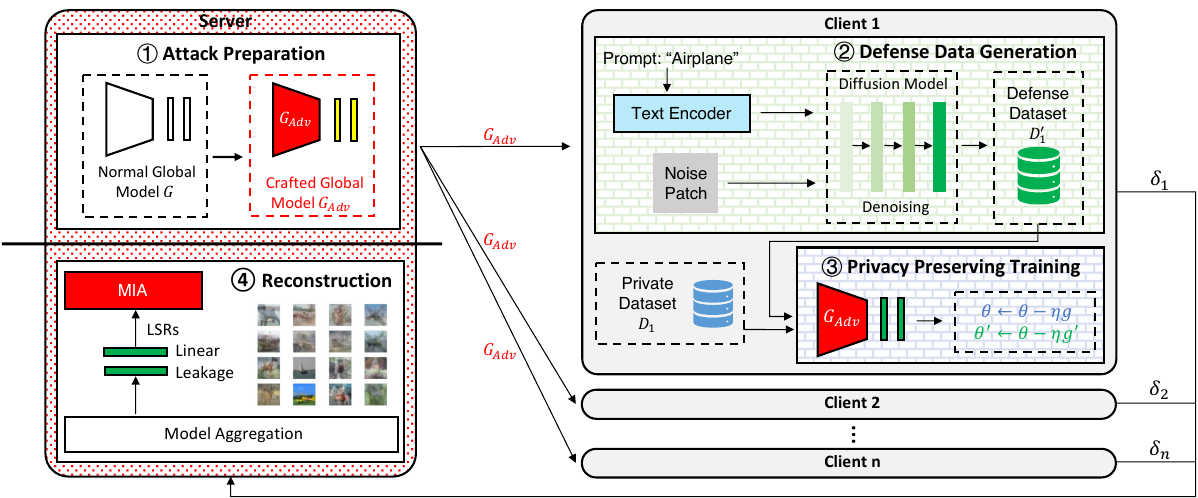}
\caption{Overview of \sysname. Each client uses a local generative model to synthesize an in-distribution defense dataset, computes a large-batch \emph{masking gradient} from it, and superimposes that masking gradient on its real model update once per local epoch. The combined update drives the effective batch size beyond the leakage capacity of the first crafted FC layer; multiple inputs collide in every leakage bin and the server's reconstruction collapses into a blended mixture rather than identifiable patient samples.}

\label{fig:system_model}
\vspace{-0cm}
\end{figure*}

\subsection{Defense Principle and Intuition}

Section~\ref{LL} established that every known MIA is bounded by the local batch size: optimization-based attacks fail at moderate batches, and linear-leakage attacks succeed only while the local batch size $B$ does not exceed the number of bins $k$ in the first crafted FC layer. We turn this universal bottleneck into a defense by deliberately driving the \emph{effective} batch size that contributes to the shared gradient above $k$, without modifying the real training data, the FL protocol, or the loss objective. Because both private and synthetic samples activate the same leakage bins, every reconstruction the server attempts is forced to collapse into a superposition of multiple inputs, and individual patient samples cannot be recovered.

In medical FL, real local datasets are typically small per institution and non-IID across institutions, so simply enlarging the real training batch is rarely possible. \sysname instead constructs an auxiliary defense dataset on the client side using a generative model, computes a separate masking gradient on it, and adds this masking gradient to the legitimate update before sharing. Because the masking gradient is computed from in-distribution synthetic data and added to the model's parameters, the masking step is itself a valid optimization step on a slightly perturbed loss landscape. The result is a structural defense that does not inject random noise (unlike DP or Outpost), does not discard informative gradient components (unlike GC or Soteria), and does not require any cryptographic protocol (unlike SA).

\subsection{Detailed Design}\label{sec:design-detail}

We describe \sysname from the perspective of one client $c_i$ in round $t$ (Fig.~\ref{fig:system_model}); all participating clients run the same procedure in parallel. Algorithm~\ref{alg:aegis} gives the complete client-side procedure.

\textbf{Step \textcircled{1}: malicious global model.} The server turns the benign global model $G$ into $G_{\mathrm{adv}}$ by inserting or re-parameterizing crafted FC layers with $k$ leakage bins, then broadcasts it to clients. From a client's view, $G_{\mathrm{adv}}$ is indistinguishable from a benign model and must be used as received.

\textbf{Step \textcircled{2}: defense dataset construction.} The client holds a local class- or text-conditional generator $\mathcal{G}$, pretrained on public data and optionally fine-tuned on a small in-domain corpus. Using local labels $\Phi_i$ (e.g., disease or organ labels), it samples $D'_i = \{\,\tilde{x}_j \sim \mathcal{G}(\,\cdot\,\mid \phi)\, : \phi \in \Phi_i,\, j = 1,\ldots,M_i\,\}$. The defense size $M_i$ is chosen so that $B + M_i > k$; in our experiments, $M_i = 2048 \gg B = 64$. The generated samples must be task-relevant: out-of-distribution samples may miss the same leakage bins as $D_i$ or bias training. We therefore filter samples by the client-side in-distribution budget $\mathbb{E}\bigl\lVert D_i - D'_i \bigr\rVert^2 \leq H$, computed over matched private and synthetic samples. Samples that violate the budget are discarded, and $\mathcal{G}$ can be fine-tuned on $D_i$ if needed. The same budget appears in Assumption~\ref{as:6}. This construction is offline, cached across rounds, and never transmitted.

\textbf{Step \textcircled{3}: privacy-preserving local training.} Each round has $E$ local epochs. In epoch $e$, the client first trains normally on mini-batches from $D_i$ with gradient $g_i^{(e)} = \nabla_{\theta} \mathcal{L}\bigl(G_{\mathrm{adv}}(D_i; \theta_i^{(e)})\bigr)$. Once per epoch, not once per round, it then computes a masking gradient on $D'_i$, $g'^{(e)}_i = \nabla_{\theta} \mathcal{L}\bigl(G_{\mathrm{adv}}(D'_i; \theta_i^{(e)})\bigr)$, using one large logical batch or gradient accumulation over micro-batches when memory is limited. The client applies the combined step $\theta_i^{(e+1)} \leftarrow \theta_i^{(e)} - \eta\, g_i^{(e)} - \eta\, g'^{(e)}_i$. Thus each epoch contributes both learning on $D_i$ and masking on $D'_i$. After $E$ epochs, the client sends $\delta_i^{t} = \theta_i^{(E)} - \theta^{t}$. The effective per-epoch direction is $\tilde{g}_i^{(e)} = (1-\lambda)\, g_i^{(e)} + \lambda\, g'^{(e)}_i$ with $\lambda = M_i / (B + M_i)$, as used in Appendix~\ref{appendix:convergence}.

\textbf{Step \textcircled{4}: server-side reconstruction collapses.} The server attacks the received $\delta_i$ (or the post-aggregation analogue under SA) as in Robbing-the-Fed~\cite{fowl2021robbing}, LOKI~\cite{zhao2023loki}, or Scale-MIA~\cite{shi2023scale}. Because masking adds $M_i$ synthetic inputs to the same leakage bins, each bin mixes multiple $\bigl(\nabla_{w_1}\mathcal{L}, \nabla_{b_1}\mathcal{L}\bigr)$ contributions. Closed-form recovery therefore returns blended images rather than identifiable patient samples.

\textbf{Why this is safe and principled.} \sysname does not perturb $D_i$, does not drop or down-weight any component of $g_i^{(e)}$, and does not modify the FL protocol; the server still receives a valid update that contains the same learning signal as standard local training plus an additional gradient step on task-relevant synthetic data. The privacy argument follows from the structural collision above, and empirical utility on deep, non-convex models is reported in Sec.~\ref{sec:experiments}.

Algorithm~\ref{alg:aegis} summarizes the complete client-side procedure; the full convergence analysis is provided in Appendix~\ref{appendix:convergence}, while Appendix~\ref{appendix:implementation} consolidates notation and implementation details.

\begin{algorithm}[t]
  \caption{\sysname client-side procedure (one communication round)}\label{alg:aegis}
  \KwIn{global model $G_{\mathrm{adv}}$ with parameters $\theta^{t}$; private data $D_i$; conditioning set $\Phi_i$; generator $\mathcal{G}$; defense size target $M_i$; in-distribution budget $H$; local hyperparameters $E, B, \eta$}
  \KwOut{client update $\delta_i^{t}$ to send to the server}
  \BlankLine
  \tcp*[l]{\textcolor{orange}{\textit{Stage 1: Defense dataset construction (offline; cached across rounds)}}}
  \If{$D'_i$ has not been constructed yet, or the modality has changed}{
    Initialize $D'_i \leftarrow \emptyset$\;
    \While{$|D'_i| < M_i$}{
      Sample a label/prompt $\phi \sim \mathrm{Uniform}(\Phi_i)$\;
      Draw a candidate sample $\tilde{x} \sim \mathcal{G}(\,\cdot\,\mid \phi)$
      \If{$\tilde{x}$ is in-distribution: $\mathbb{E}\lVert D_i - (D'_i \cup \{\tilde{x}\})\rVert^2 \leq H$}{
        $D'_i \leftarrow D'_i \cup \{\tilde{x}\}$\;
      }
      \Else{
        Discard $\tilde{x}$; if rejection rate is high, fine-tune $\mathcal{G}$ on a sanitized subset of $D_i$ and resume sampling\;
      }
    }
    Cache $D'_i$ locally; reuse in subsequent rounds\;
  }
  \BlankLine
  \tcp*[l]{\textcolor{orange}{\textit{Stage 2: Local training interleaved with per-epoch masking}}}
  $\theta_i \leftarrow \theta^{t}$ \tcp*[r]{initialize local copy from received global model}
  \For{epoch $e = 1, \ldots, E$}{
    Partition $D_i$ into mini-batches of size $B$\;
    \For{each mini-batch $B_j \subseteq D_i$}{
      $g_{i,j} \leftarrow \nabla_{\theta} \mathcal{L}\bigl(G_{\mathrm{adv}}(B_j; \theta_i)\bigr)$\;
      $\theta_i \leftarrow \theta_i - \eta\, g_{i,j}$ \tcp*[r]{standard SGD step}
    }
    \BlankLine
    \tcp*[l]{\textcolor{orange}{\textit{Per-epoch masking: once per local epoch, not once per round}}}
    Form a large batch from $D'_i$ of size $M_i$ (or accumulate gradients over micro-batches when memory-limited)\;
    $g'^{(e)}_i \leftarrow \nabla_{\theta} \mathcal{L}\bigl(G_{\mathrm{adv}}(D'_i; \theta_i)\bigr)$\;
    $\theta_i \leftarrow \theta_i - \eta\, g'^{(e)}_i$ \tcp*[r]{drives effective batch size $> k$}
  }
  \BlankLine
  \tcp*[l]{\textcolor{orange}{\textit{Stage 3: Form and transmit the masked update}}}
  $\delta_i^{t} \leftarrow \theta_i - \theta^{t}$ \tcp*[r]{accumulated parameter delta after all $E$ epochs of interleaved learning and masking}
  Send $\delta_i^{t}$ to the server (through SA or directly, per the FL protocol)\;
\end{algorithm}

\section{Experiments}\label{sec:experiments}

We evaluate \sysname as a privacy-preserving primitive for medical FL, focusing on two questions: (i) does \sysname suppress state-of-the-art linear-leakage MIAs on both standard benchmarks and clinical imaging modalities, and (ii) does it preserve global model utility competitively with state-of-the-art defenses, without imposing prohibitive overhead?

\subsection{Experimental Setup}\label{sec:exp-setup}

\textbf{Attack Baselines:} To evaluate the effectiveness of our mechanism, we select three representative closed-form attacks: Robbing-the-Fed~\cite{fowl2021robbing}, LOKI~\cite{zhao2023loki}, Scale-MIA~\cite{shi2023scale}. These are evaluated in a production FL setting where optimization-based methods fail beyond their feasibility.

\textbf{Defense Baselines:} We compare the effectiveness of \sysname with various state-of-the-art defense schemes. Among them, DP~\cite{46432} adds artificial noise to the local training gradients. Outpost~\cite{wang2023more} and GradDefense (GD)~\cite{wang2022protect} adaptively add noise to gradients based on privacy leakage risk. Soteria~\cite{sun2021soteria} and Gradient Compression (GC)~\cite{zhu2019deep} prune parts of the gradients to hide the recoverable ground truth.

\textbf{FL configuration.} We use non-IID FedAvg with $N=100$ clients, disjoint local data, $10\%$ client participation per round, $E=3$ local epochs, and batch size $B=64$.

\textbf{Datasets.} We use MNIST and CIFAR-10 for standard utility/privacy comparisons, and ChestMNIST, OrganAMNIST, and PathMNIST from MedMNIST v2~\cite{yang2023medmnist} for medical FL evaluation.

\textbf{Defense data.} Each client builds $D'_i$ with a class- or text-conditional latent diffusion model~\cite{Rombach_2022_CVPR}; generated samples are filtered by $\mathbb{E}\lVert D_i - D'_i\rVert^2 \leq H$ and used only for masking gradients. Unless swept, \sysname uses defense batch size $M_i=2048$.

\textbf{Evaluation Metrics:} We assess reconstruction performance using peak signal-to-noise ratio (PSNR) and structural similarity index measure (SSIM). PSNR quantifies the ratio of an image signal's maximum possible power to noise, measured in decibels (dB). SSIM evaluates perceived image quality by comparing luminance, contrast, and structure, with values ranging from -1 to 1, where scores near 1 indicate high similarity and those below 0.5 suggest significant differences. The reconstruction rate is defined as the fraction of reconstructed images with PSNR above the threshold $th=18$. These metrics are widely adopted in prior work~\cite{shi2023scale, fowl2021robbing, wen2022fishing, zhao2023loki}. To assess \sysname’s impact on FL performance, we track test accuracy across FL rounds and compare it against other defense baselines.

\textit{Details.} Appendix~\ref{appendix:exp_details} gives model architectures, optimization, data partitioning, generation prompts/filtering, baseline hyperparameters, and attack protocol; Appendix~\ref{appendix:discussion} states trust assumptions and limitations.

\subsection{Round-to-Accuracy under Defense}

We first evaluate how \sysname affects the global model's convergence trajectory. We set the \sysname defense batch size to 2048; for baselines, the Soteria pruning rate is 80\%, the GC pruning rate is 80\%, DP uses Gaussian noise with $\varepsilon = 1.0$ and $\Delta = 10^{-4}$, and GD/Outpost use the configurations recommended in their original papers~\cite{wang2022protect, wang2023more}.

\begin{figure}[t]
\centering
\includegraphics[width=0.8\linewidth]{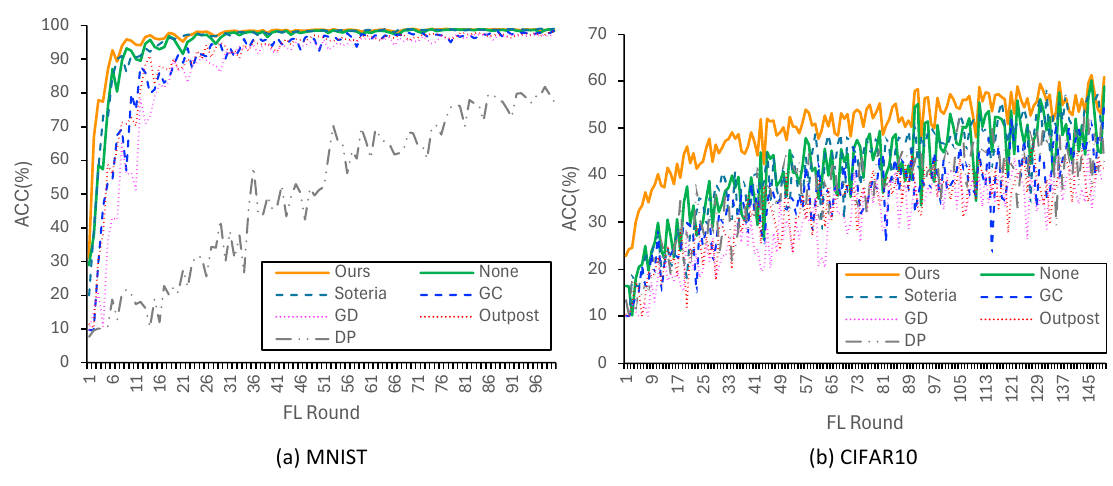}
\caption{Round-to-accuracy comparison of \sysname (defense batch size 2048) against state-of-the-art defenses on (a) MNIST and (b) CIFAR-10.}
\label{fig:ACC}
\end{figure}

On MNIST (Fig.~\ref{fig:ACC}(a)), DP is the most damaging baseline: its calibrated noise visibly slows convergence and lowers final accuracy. Adaptive-noise schemes (GD, Outpost) attenuate this effect by injecting smaller, risk-calibrated perturbations. Pruning-based defenses (GC, Soteria) only remove small-magnitude gradients and therefore incur smaller utility hits. \sysname tracks the no-defense baseline most closely and achieves the highest final test accuracy among all defenses. CIFAR-10 (Fig.~\ref{fig:ACC}(b)) shows the same ordering: \sysname's masking gradients are valid optimization steps on in-distribution data, and therefore do not destabilize convergence the way noise- or pruning-based defenses do. The convergence bound in Appendix~\ref{appendix:convergence} is stated under standard convex assumptions and explains how the auxiliary-data discrepancy enters constants; it is not a non-convex guarantee for the deep networks here, so Fig.~\ref{fig:ACC} is the primary utility evidence in the deep regime.

\begin{table}[t]
    \centering
    \caption{Wall-clock training time (seconds) for 100 FL rounds on MNIST and 150 FL rounds on CIFAR-10.}
    \label{tab:running_time}
    \vspace{-2mm}
    \begin{tabular}{lrrrrrrr}
        \toprule
        Dataset & None & \sysname{} & Soteria & GC & GD & Outpost & DP \\
        \midrule
        MNIST    & 290.5 & 1890.1 & 19816.2 & 334.5 & 4622.4 & 512.4 & 330.9 \\
        CIFAR-10 & 433.1 & 2339.3 & 32482.0 & 512.1 & 8356.8 & 781.7 & 517.0 \\
        \bottomrule
    \end{tabular}
\end{table}

Tab.~\ref{tab:running_time} reports wall-clock training times. Noise- and pruning-based defenses (DP, Outpost, GC) introduce minimal overhead. GD and Soteria, by contrast, require additional per-epoch computations: Soteria learns perturbed representations every epoch, and GD computes a model-sensitivity metric to calibrate Gaussian noise. \sysname adds only the cost of one large-batch forward-backward pass over $D'_i$ per local epoch and remains substantially faster than GD and Soteria, while offering substantially stronger privacy protection (Sec.~\ref{EE}). When GPU memory is smaller than the logical masking batch $M_i$, the client can accumulate gradients over micro-batches of size $b$: the mean gradient matches a single full-batch pass, so bin collisions and the privacy argument are unchanged, while wall-clock grows roughly with $M_i/b$. This overhead is modest on our workstation but can matter for resource-constrained clients (Appendix~\ref{appendix:discussion}).

\subsection{Effectiveness against Linear-Leakage MIAs}
\label{EE}

\begin{figure*}[h!]
\centering
\vspace{-0cm}
\includegraphics[width=0.99\linewidth]{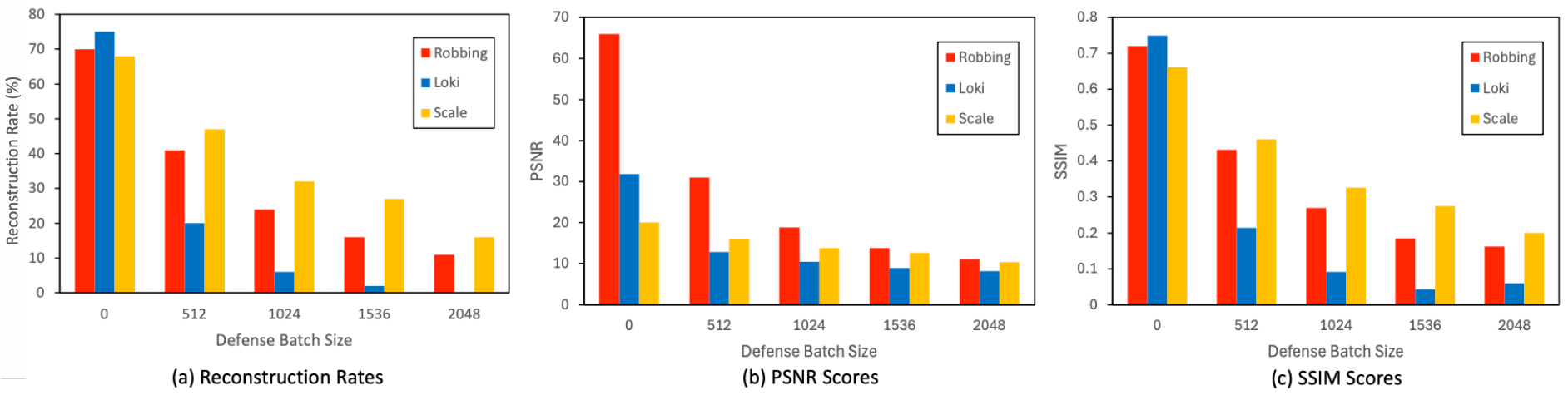}
\caption{Effectiveness of \sysname against three state-of-the-art linear-leakage MIAs (Robbing-the-Fed, LOKI, Scale-MIA) on CIFAR-10. As the defense batch size grows, RR, PSNR, and SSIM all decrease consistently.}
\label{fig:D1}
\vspace{-0cm}
\end{figure*}

\begin{table*}[h!]
    \centering
    \caption{Robbing-the-Fed reconstructions on three MedMNIST modalities (ChestMNIST, OrganAMNIST, PathMNIST) under no defense, \sysname, DP, and GC. Each block shows qualitative samples (top) and quantitative reconstruction metrics (bottom). Blue scores in parentheses are computed only over successfully reconstructed samples (PSNR$>$18); RR is the overall reconstruction rate (fraction of samples with PSNR$>$18).}

    \label{tab:recover_medmnist}
    \setlength{\tabcolsep}{4pt}
    \renewcommand{\arraystretch}{1.12}
    \begin{tabular}{c@{\hspace{1.2em}}cccc}
        \hline
        \textbf{Ground Truth} & \textbf{No defense} & \textbf{\sysname} & \textbf{DP} & \textbf{GC} \\
        \hline
        \multicolumn{5}{c}{\scriptsize ChestMNIST (chest X-ray)} \\
        \hline
        \includegraphics[width=0.17\linewidth]{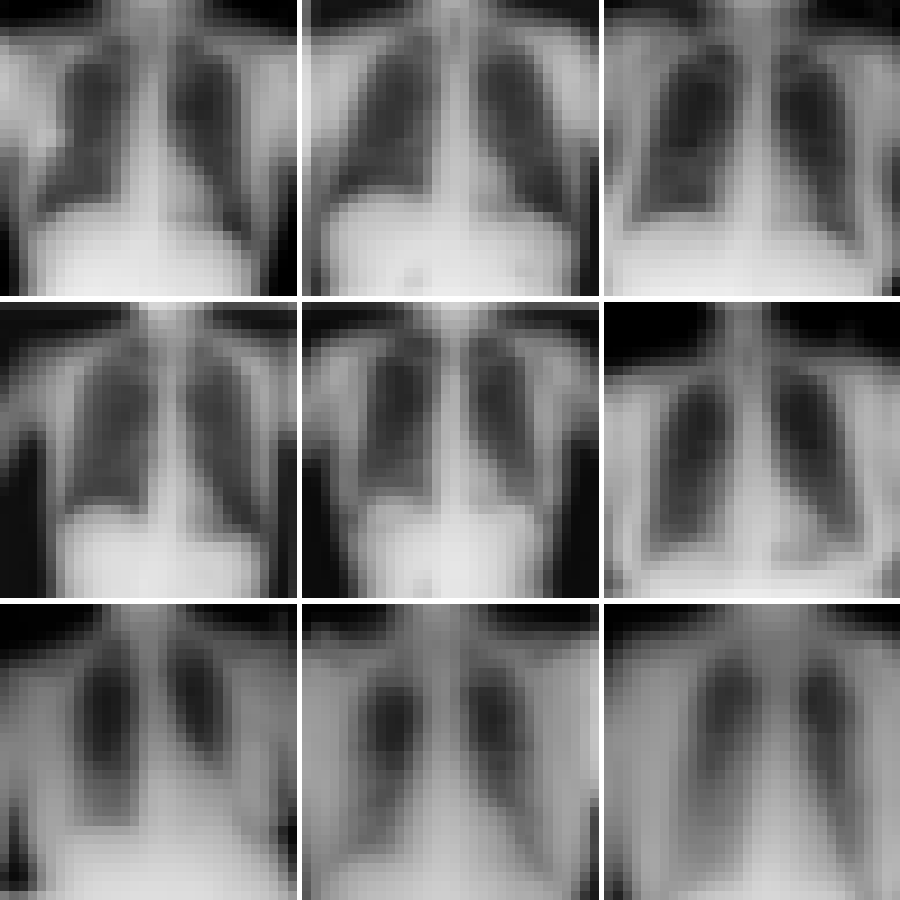} &
        \includegraphics[width=0.17\linewidth]{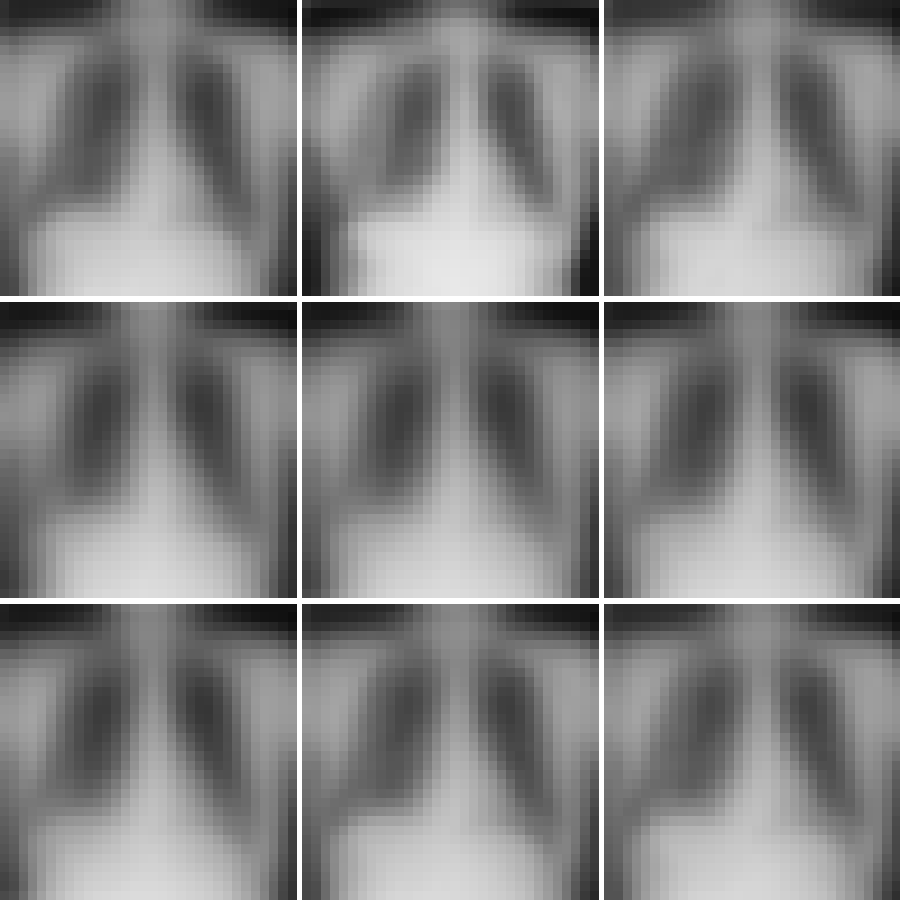} &
        \includegraphics[width=0.17\linewidth]{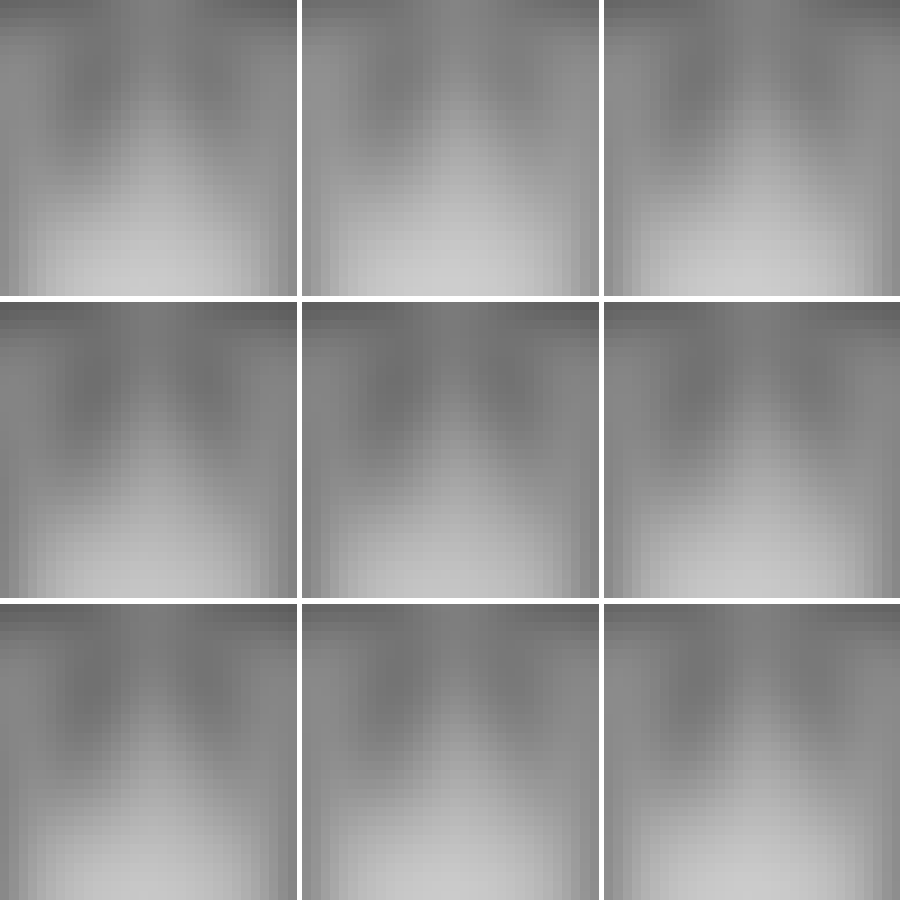} &
        \includegraphics[width=0.17\linewidth]{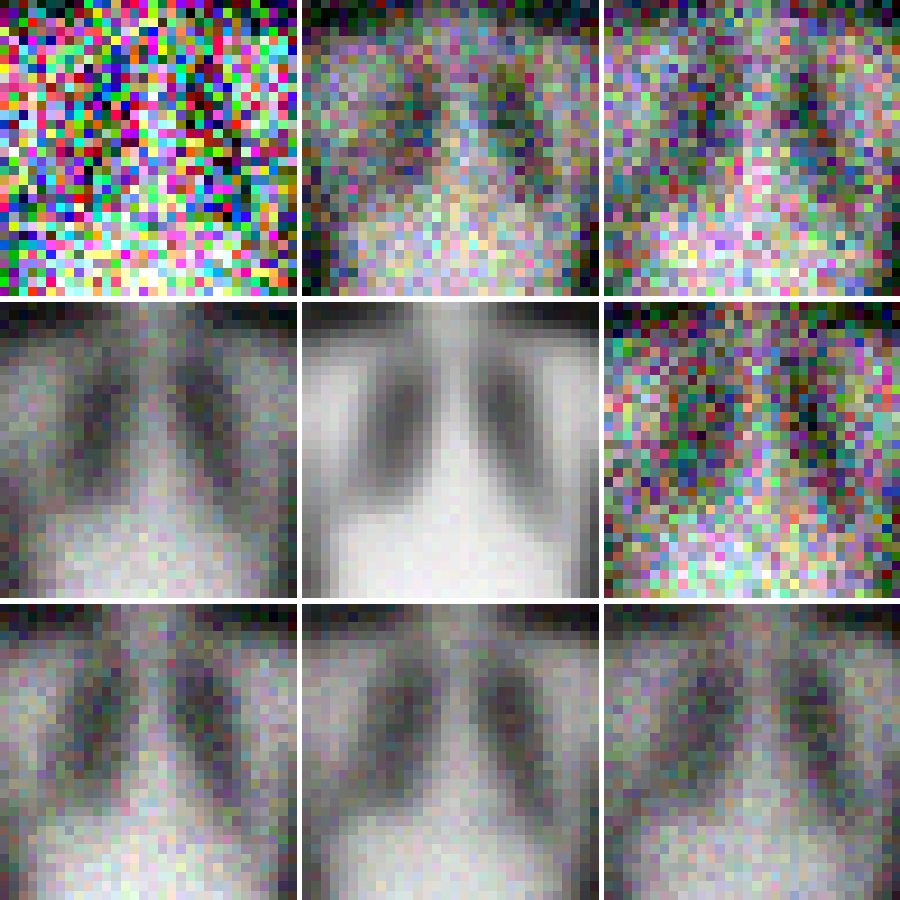} &
        \includegraphics[width=0.17\linewidth]{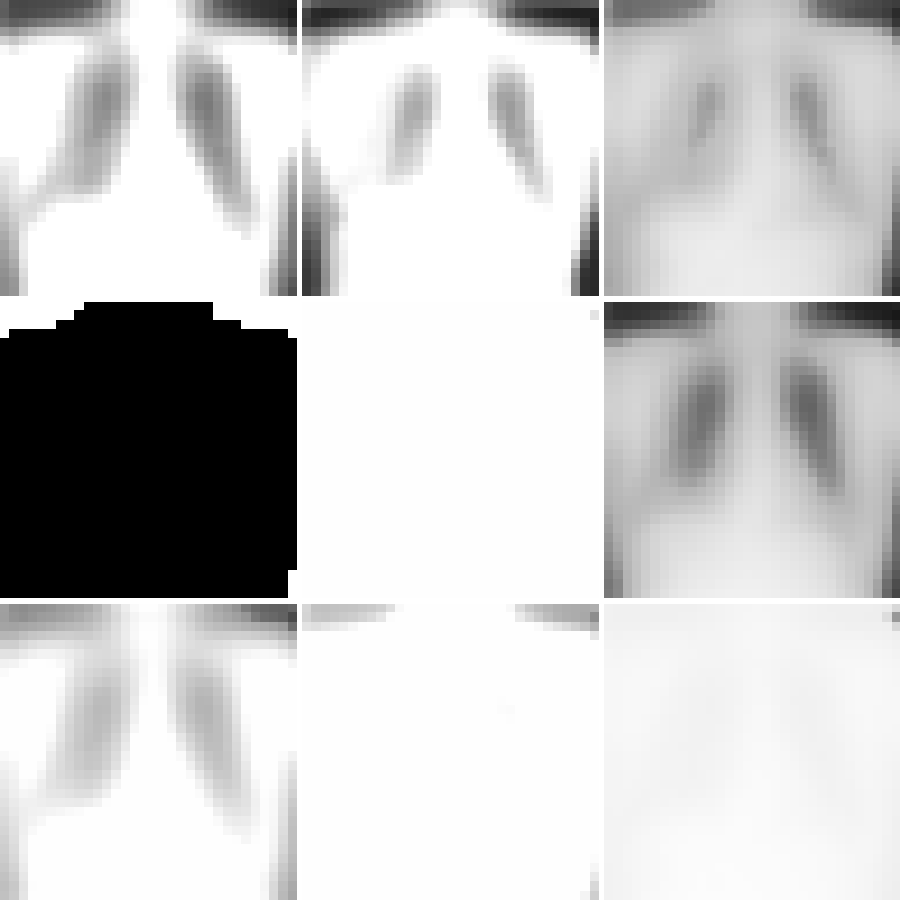} \\
        \hline
        \textbf{RR (\%)} & 89.06 & 9.50 & 71.88 & 59.38 \\
        \hline
        \textbf{PSNR} & 62.40 {\scriptsize\textcolor{blue}{(105.40)}} & 18.70 {\scriptsize\textcolor{blue}{(29.80)}} & 36.80 {\scriptsize\textcolor{blue}{(58.60)}} & 42.50 {\scriptsize\textcolor{blue}{(76.20)}} \\
        \hline
        \textbf{SSIM} & 0.94 {\scriptsize\textcolor{blue}{(0.99)}} & 0.46 {\scriptsize\textcolor{blue}{(0.91)}} & 0.82 {\scriptsize\textcolor{blue}{(0.96)}} & 0.78 {\scriptsize\textcolor{blue}{(0.97)}} \\
        \hline\hline

        \multicolumn{5}{c}{\scriptsize OrganAMNIST (abdominal CT)} \\
        \hline
        \includegraphics[width=0.17\linewidth]{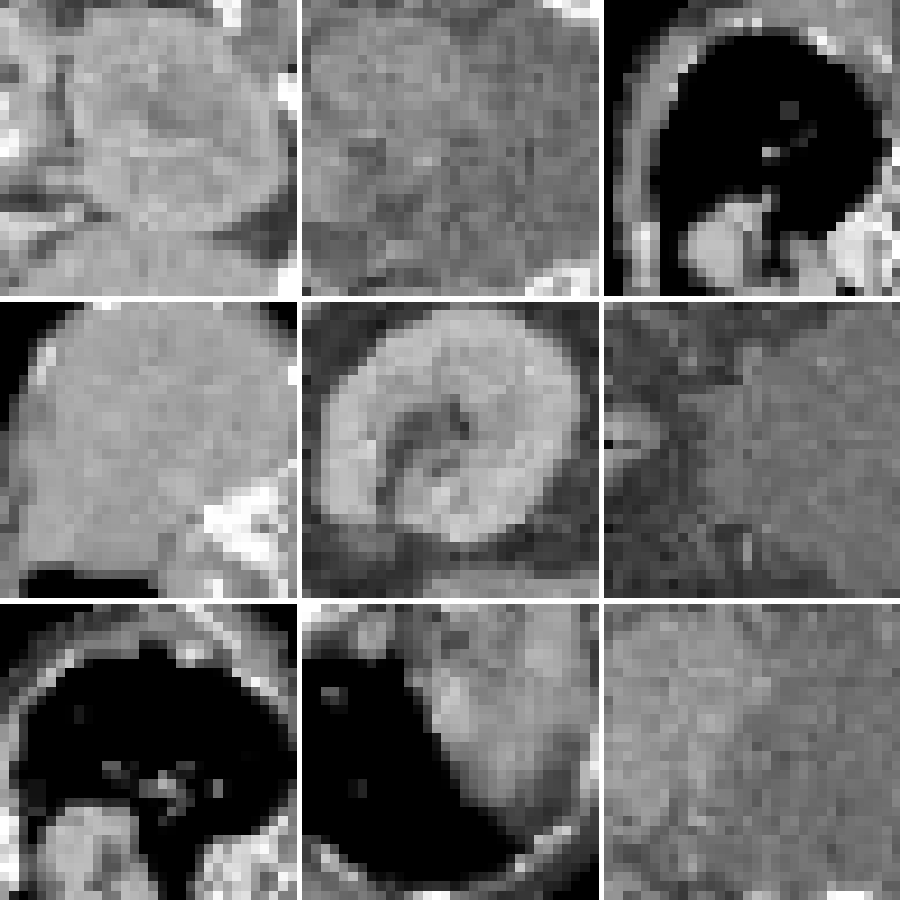} &
        \includegraphics[width=0.17\linewidth]{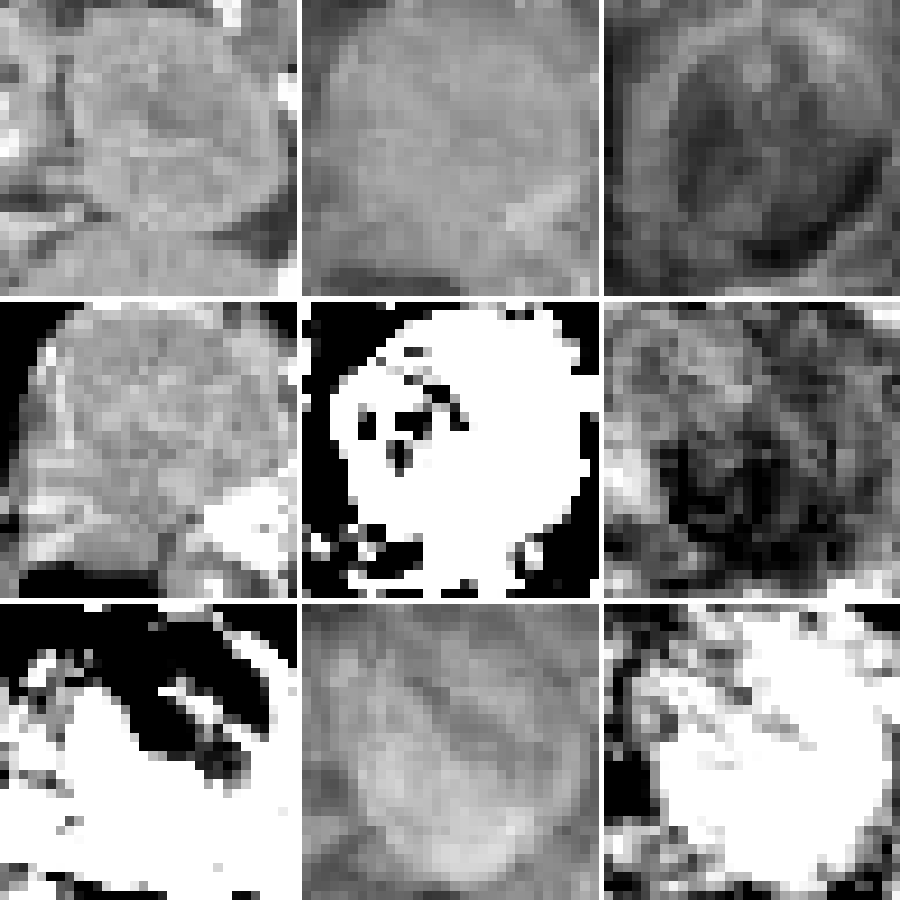} &
        \includegraphics[width=0.17\linewidth]{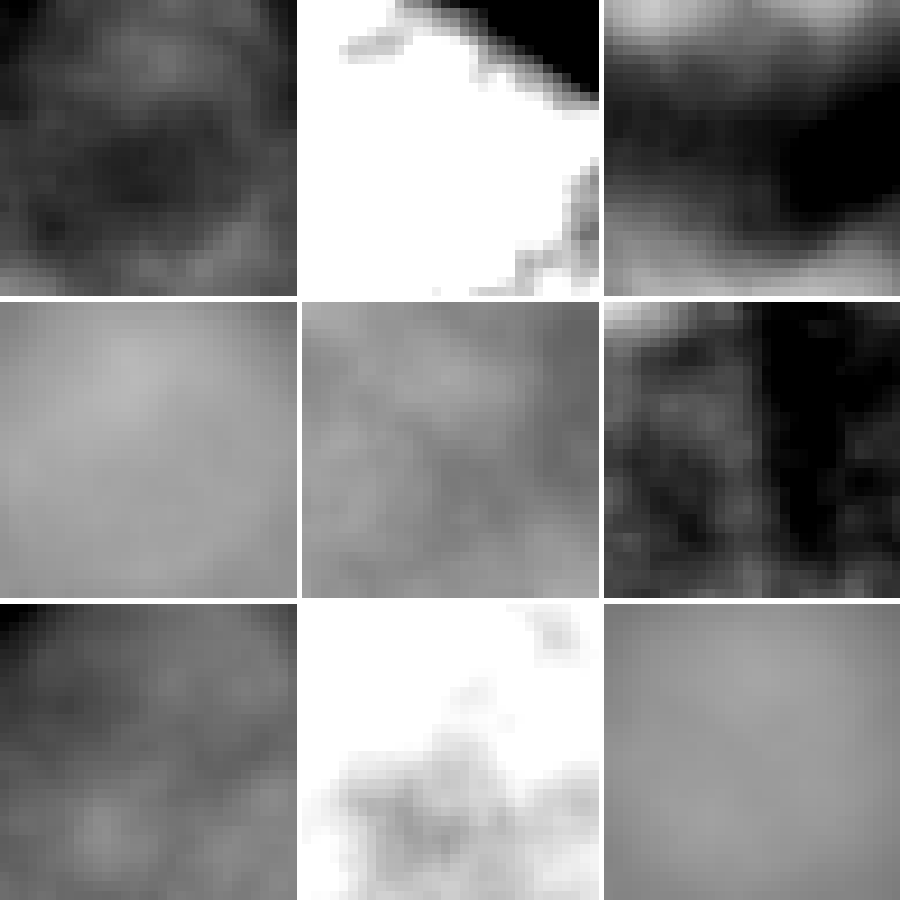} &
        \includegraphics[width=0.17\linewidth]{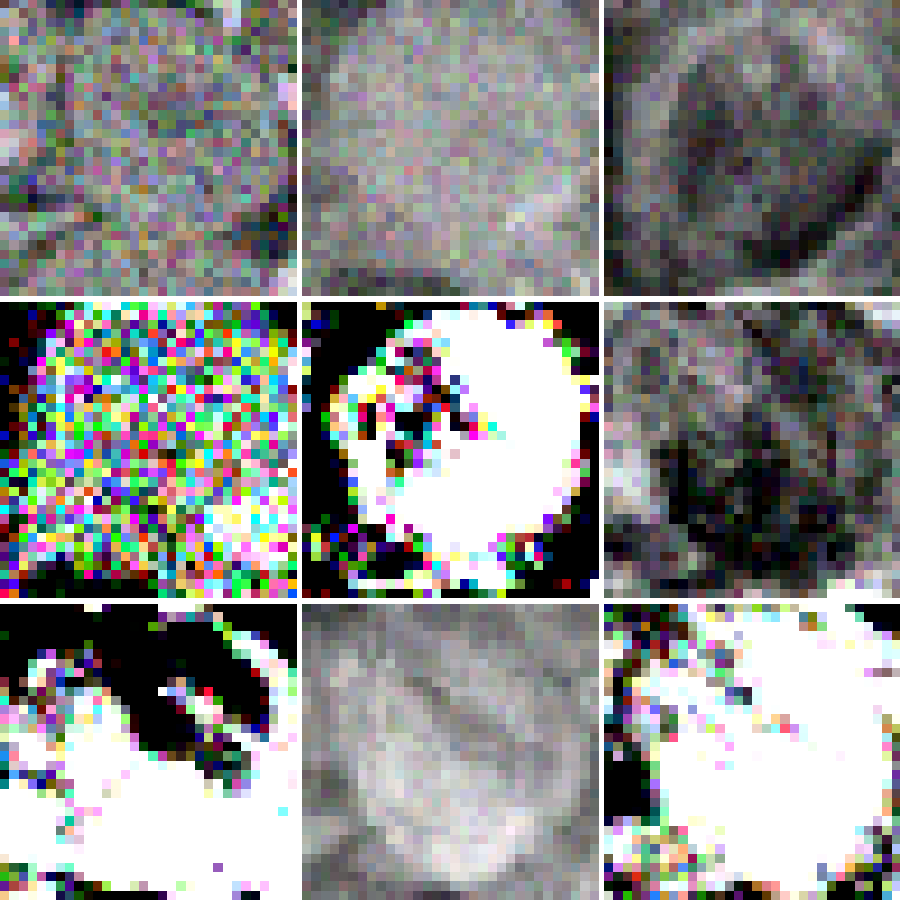} &
        \includegraphics[width=0.17\linewidth]{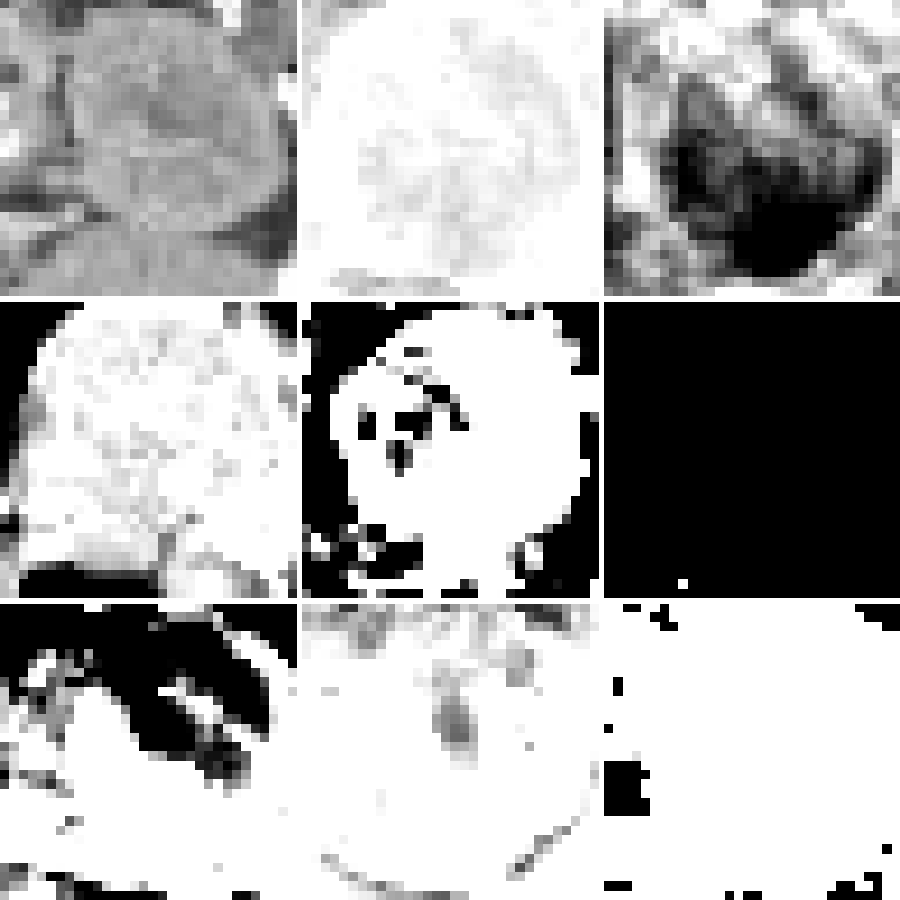} \\
        \hline
        \textbf{RR (\%)} & 82.81 & 11.63 & 65.63 & 53.13 \\
        \hline
        \textbf{PSNR} & 50.60 {\scriptsize\textcolor{blue}{(98.70)}} & 20.10 {\scriptsize\textcolor{blue}{(31.40)}} & 33.40 {\scriptsize\textcolor{blue}{(54.20)}} & 38.90 {\scriptsize\textcolor{blue}{(70.50)}} \\
        \hline
        \textbf{SSIM} & 0.89 {\scriptsize\textcolor{blue}{(0.98)}} & 0.51 {\scriptsize\textcolor{blue}{(0.88)}} & 0.76 {\scriptsize\textcolor{blue}{(0.94)}} & 0.72 {\scriptsize\textcolor{blue}{(0.95)}} \\
        \hline\hline

        \multicolumn{5}{c}{\scriptsize PathMNIST (colon pathology)} \\
        \hline
        \includegraphics[width=0.17\linewidth]{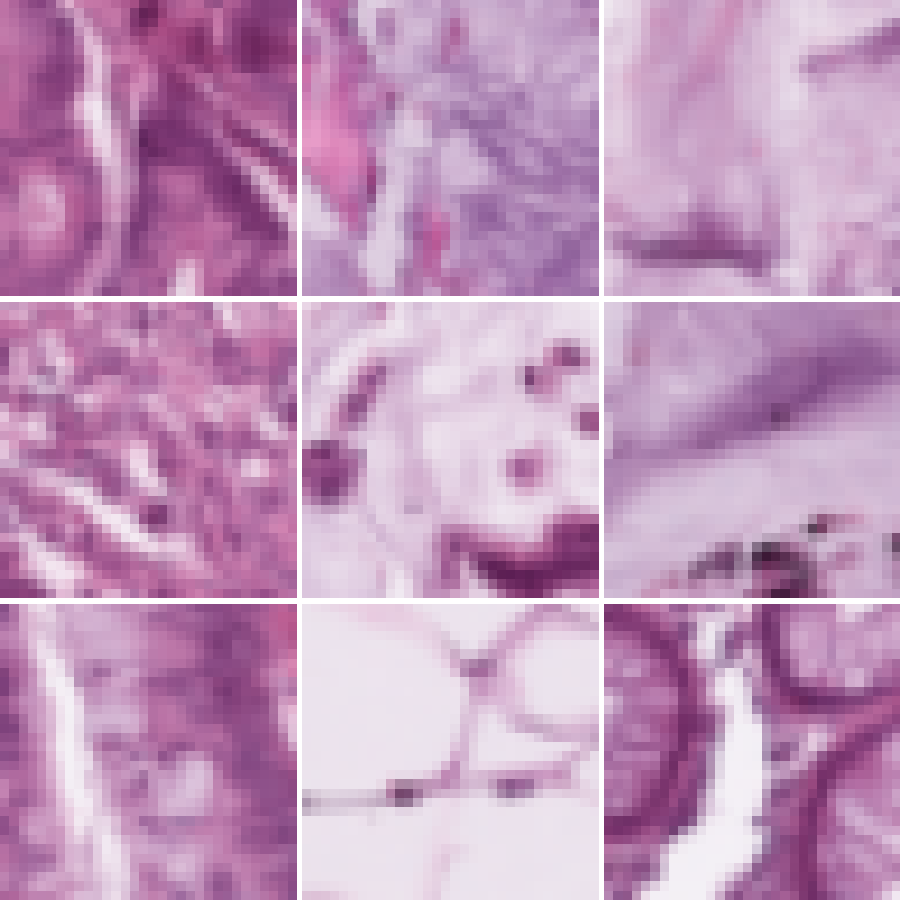} &
        \includegraphics[width=0.17\linewidth]{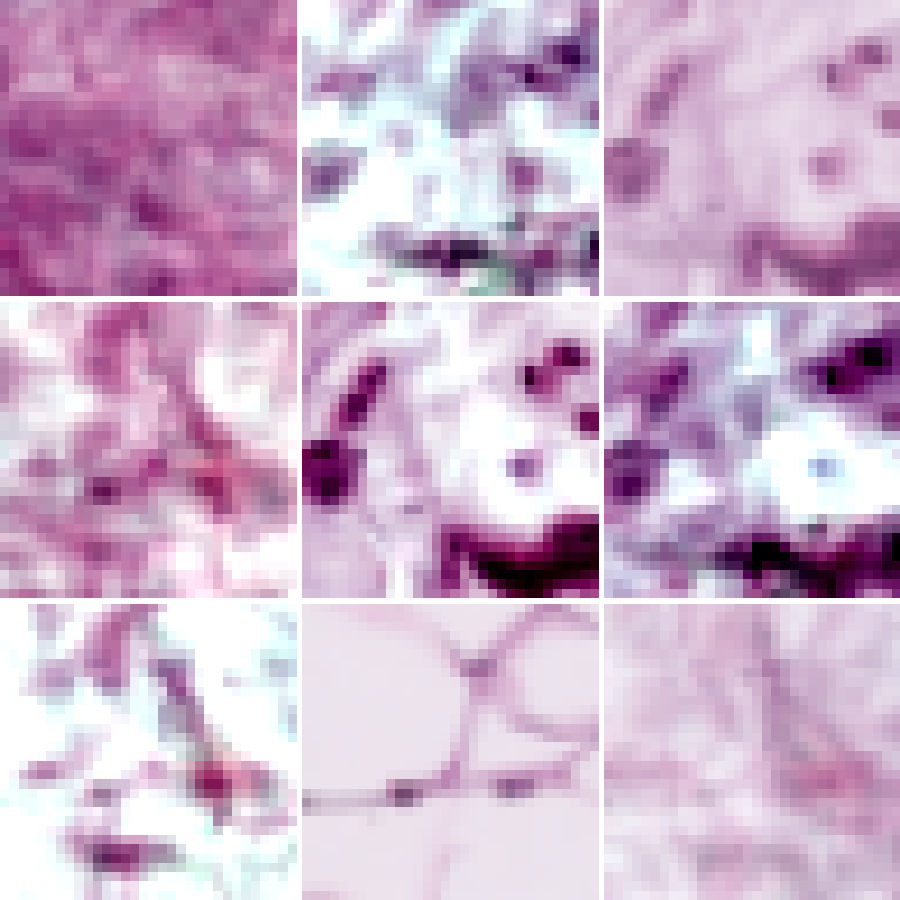} &
        \includegraphics[width=0.17\linewidth]{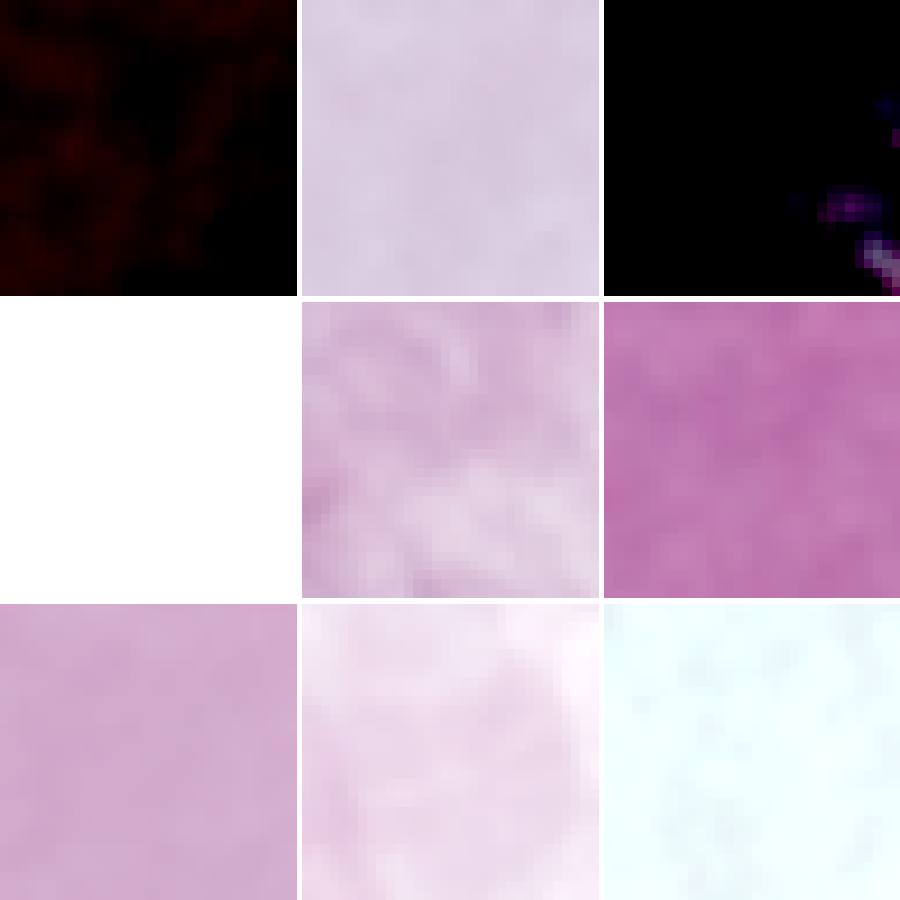} &
        \includegraphics[width=0.17\linewidth]{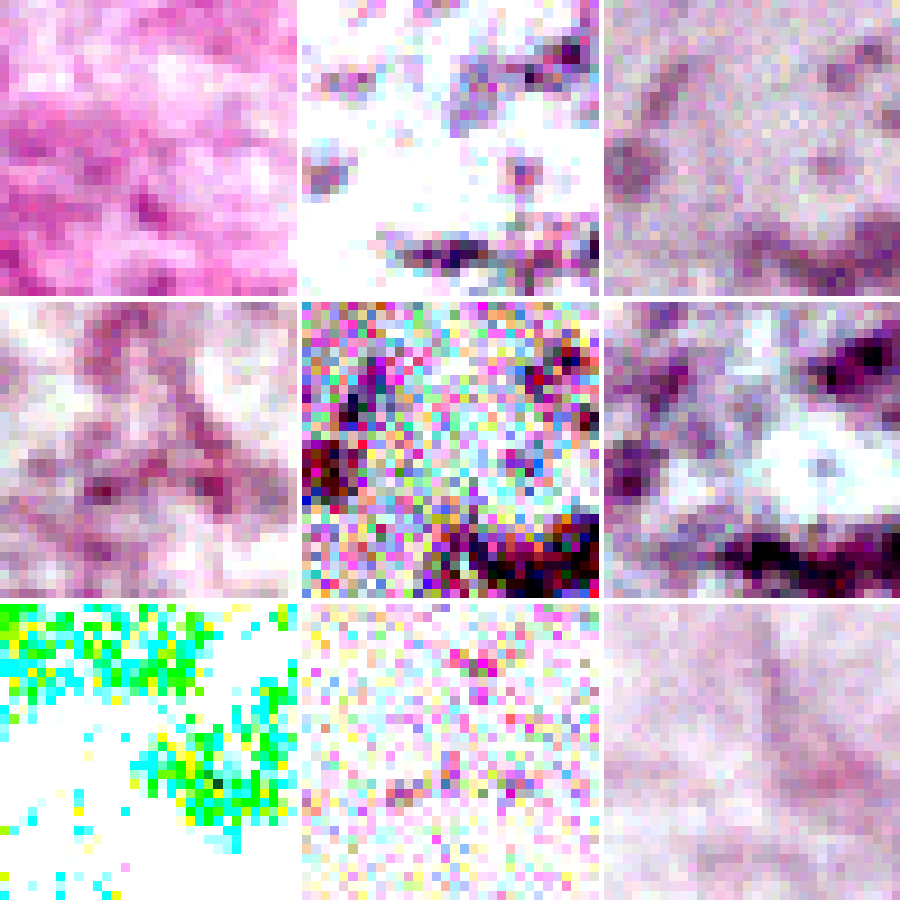} &
        \includegraphics[width=0.17\linewidth]{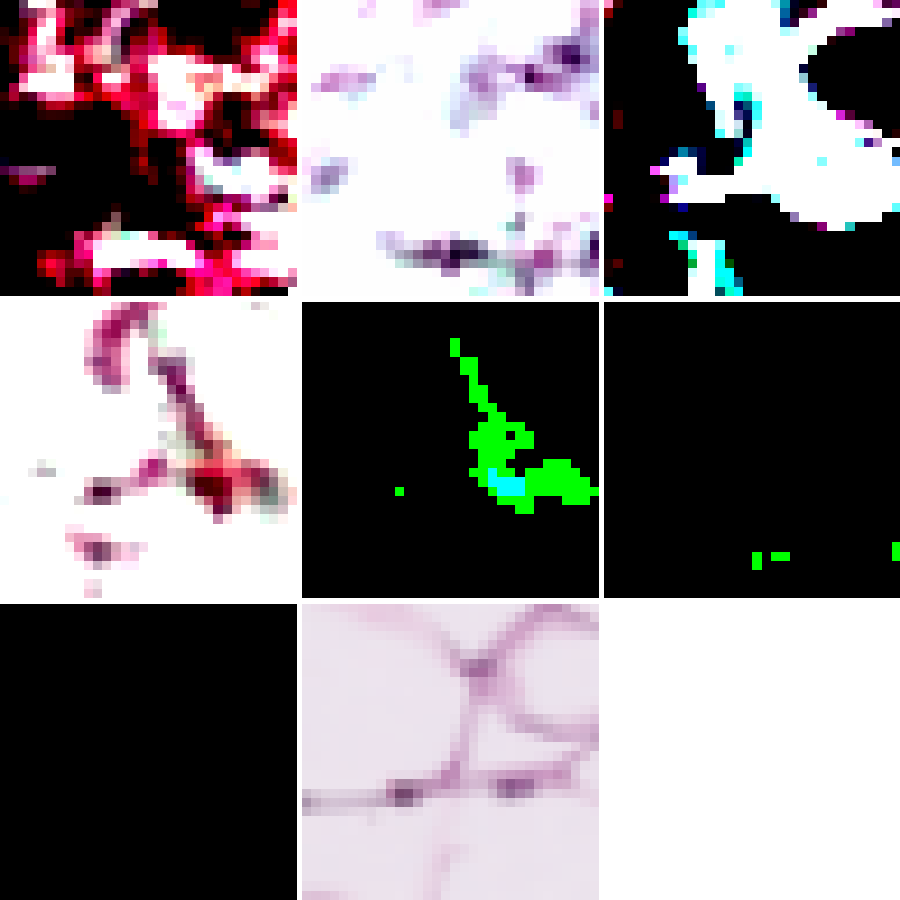} \\
        \hline
        \textbf{RR (\%)} & 78.13 & 9.38 & 59.38 & 46.88 \\
        \hline
        \textbf{PSNR} & 45.20 {\scriptsize\textcolor{blue}{(91.30)}} & 17.30 {\scriptsize\textcolor{blue}{(27.90)}} & 30.60 {\scriptsize\textcolor{blue}{(48.70)}} & 34.20 {\scriptsize\textcolor{blue}{(61.80)}} \\
        \hline
        \textbf{SSIM} & 0.86 {\scriptsize\textcolor{blue}{(0.97)}} & 0.38 {\scriptsize\textcolor{blue}{(0.84)}} & 0.69 {\scriptsize\textcolor{blue}{(0.92)}} & 0.64 {\scriptsize\textcolor{blue}{(0.93)}} \\
        \hline
    \end{tabular}
    \vspace{-0cm}
\end{table*}

We now evaluate \sysname's defensive effectiveness against the three linear-leakage MIAs. Experiments use the same FL setup ($E=3$, $B=64$) for direct comparability. Fig.~\ref{fig:D1} shows that, across all three attacks, \sysname systematically suppresses reconstruction quality: as the defense batch size grows, RR, PSNR, and SSIM all decrease together, and reconstructed samples become progressively indistinguishable from noise. This empirically confirms our central claim that enlarging the effective batch size beyond the linear-leakage capacity, via a synthetic-data masking gradient, is sufficient to disrupt scalable, closed-form reconstructions. Quantitative summaries and additional reconstructed samples comparing \sysname against DP, GC, Soteria, and Outpost on CIFAR-10 are provided in Appendix~\ref{appendix:cifar_robbing_additional}. The experiments instantiate the standard linear-leakage construction; if a server increases the first-FC width $k$, \sysname correspondingly requires a larger $M_i$, so sizing the defense batch under unknown or adaptive $k$ is a deployment consideration rather than a protocol change. A server might also attempt to demix bins using public knowledge of the generator (typically ill-posed when real and synthetic contributions mix), or pursue primitives outside linear leakage (not among today's scalable MIAs). Appendix~\ref{appendix:discussion} expands these adaptive-adversary considerations.

Tab.~\ref{tab:recover_medmnist} reports Robbing-the-Fed reconstructions on the three MedMNIST modalities with both qualitative samples and quantitative RR/PSNR/SSIM (computed as in Tab.~\ref{tab:recover-cifar-appendix}), providing the most safety-relevant evidence: anatomical and pathological detail in real chest X-rays, abdominal CTs, and colon pathology tiles. Without any defense, RR reaches $78$--$89\%$ with high PSNR/SSIM; DP lowers RR but leaves strong conditional scores on surviving reconstructions; GC yields intermediate RR with similarly high PSNR on recovered tiles when pruning misses informative gradients. \sysname drives RR down to $9.38$--$11.63\%$ across modalities while collapsing aggregate PSNR and SSIM relative to no defense and both baselines, matching the blended visual outputs: gradient superposition forces multiple medical samples into the same leakage bins and suppresses identifiable anatomical or pathological detail. Crucially, this protection is achieved without perturbing or discarding any informative gradient component; the model still learns from the full, untouched real medical data, so model utility (Tab.~\ref{tab:medical-accuracy}) is preserved while privacy is restored. These medical studies use three MedMNIST modalities at benchmark resolution; extending to full-resolution CT volumes or whole-slide pathology would require correspondingly stronger generators and is outside the experiments reported here. The quantitative claims target FC-layer linear leakage as in Robbing-the-Fed-class attacks; defenses against hypothetical primitives that bypass this bottleneck are outside our scope (Appendix~\ref{appendix:discussion}).

\section{Conclusion}

We presented \sysname, a principled, protocol-compatible defense against linear-leakage model inversion attacks in medical federated learning. By identifying that all known scalable MIAs are bounded by the local batch size relative to the first-FC-layer capacity, \sysname turns this structural bottleneck into a defense: each client computes a large-batch masking gradient on locally synthesized task-relevant data and superimposes it onto its real update once per local epoch, forcing the server's closed-form reconstructions to collapse into indistinguishable mixtures. \sysname is not tied to a specific generative architecture; it requires only that the auxiliary samples are sufficiently task-relevant to avoid harming utility while inducing gradient collisions. We complement the design with stability analysis under standard convex assumptions (not a non-convex guarantee) and show, on MNIST, CIFAR-10, and three MedMNIST modalities (chest X-ray, abdominal CT, colon pathology), that \sysname suppresses three state-of-the-art MIAs while preserving model utility and incurring competitive overhead, providing a practical privacy-preserving primitive for AI/ML for health and biotechnology.

\section*{Acknowledgements}

This work was supported in part by the Office of Naval Research under grants N00014-24-1-2730, and the National Science Foundation under grants 2312447, 2247560, and 2235232.

\clearpage
{
    \small
    \bibliographystyle{plainnat}
    \bibliography{main}
}

\clearpage
\appendix

\setcounter{table}{0}
\renewcommand{\thetable}{A.\arabic{table}}
\setcounter{figure}{0}
\renewcommand{\thefigure}{A.\arabic{figure}}
\setcounter{algocf}{0}
\renewcommand{\thealgocf}{A.\arabic{algocf}}
\setcounter{equation}{0}
\renewcommand{\theequation}{A.\arabic{equation}}

\section*{Appendix}
The appendix is organized as follows. Appendix~\ref{appendix:implementation} gives a consolidated notation table and implementation details for Algorithm~\ref{alg:aegis}. Appendix~\ref{appendix:exp_details} records the experimental setup needed for reproducibility, including data partitioning, models, optimization, defense generation, baseline settings, and attack protocol. Appendix~\ref{appendix:local_robbing} reports the local-hyperparameter sensitivity study referenced from Sec.~\ref{sec:threat-model}. Appendix~\ref{appendix:cifar_robbing_additional} provides additional CIFAR-10 reconstruction comparisons against five defense baselines. Appendix~\ref{appendix:convergence} gives the standard convex assumptions and convergence theorem used in~\cite{Li2020On}, adapted here. Appendix~\ref{appendix:mia_survey} summarizes representative gradient-based MIAs in FL. Appendix~\ref{appendix:discussion} collects extended discussion of deployment assumptions and scope.

Additional medical utility and privacy--utility results appear in Appendix~\ref{appendix:medical}; generator, non-IID, and backbone experiments appear in Appendix~\ref{appendix:sensitivity}.

\section{\sysname Notation and Implementation Details}
\label{appendix:implementation}

This appendix supplements the client-side procedure in Algorithm~\ref{alg:aegis}. The same procedure runs in parallel on every participating client, and the server-side aggregation step is unchanged from the underlying FL protocol (FedAvg or FedSGD with or without secure aggregation).

\subsection{Notation and Inputs}

Table~\ref{tab:notation} consolidates the symbols used in the client-side algorithm and the convergence analysis.

\begin{table}[h!]
    \centering
    \small
    \caption{Notation used by \sysname. The left column lists the mathematical symbol and the right column gives its role in the FL protocol, defense mechanism, or convergence analysis.}
    \label{tab:notation}
    \begin{tabular}{lp{0.68\linewidth}}
        \hline
        \textbf{Notation} & \textbf{Explanation} \\
        \hline
        $c_i$, $N$ & Client $i$ and the total number of clients. \\
        $t$, $E$, $B$ & Communication round, number of local epochs, and local mini-batch size. \\
        $K$, $S_t$ & Number and set of clients selected in round $t$. \\
        $G$, $G_{\mathrm{adv}}$ & Benign global model and potentially adversarial broadcast model. \\
        $\theta^t$, $\theta_i$ & Global parameters at round $t$ and the local copy maintained by client $c_i$. \\
        $\delta_i^t$ & Client update sent to the server after local training and masking. \\
        $D_i$, $D'_i$ & Private client dataset and synthetic defense dataset. \\
        $\Phi_i$, $\mathcal{G}$ & Local label/prompt set and class- or text-conditional generator. \\
        $M_i$, $k$ & Defense dataset size and leakage-bin capacity of the first crafted FC layer. \\
        $H$ & In-distribution budget, $\mathbb{E}\lVert D_i-D'_i\rVert^2\leq H$. \\
        $\eta$, $g_{i,j}$ & Local learning rate and mini-batch gradient on private data. \\
        $g_i^{(e)}$, $g_i'^{(e)}$ & Epoch-level private gradient and masking gradient. \\
        $\lambda$ & Mixing weight $M_i/(B+M_i)$ used in the effective-gradient view. \\
        $F_i$, $F$, $p_i$ & Local objective, global objective, and client aggregation weight. \\
        $L$, $\mu$, $\sigma_i^2$, $G^2$ & Smoothness, strong-convexity, gradient-variance, and gradient-norm constants. \\
        $\kappa$, $\gamma$, $\Gamma$ & Condition number, learning-rate schedule constant, and client heterogeneity term. \\
        \hline
    \end{tabular}
\end{table}

\subsection{Implementation Notes}

\textbf{Cost amortization.} Stage~1 is performed once and cached, so \sysname adds one masking pass over $D'_i$ (size $M_i$) per local epoch, i.e., $E$ extra forward-backward passes per communication round. This is empirically much cheaper than the per-epoch costs incurred by GD and Soteria (Tab.~\ref{tab:running_time}).

\textbf{Memory considerations and gradient accumulation.} The masking step requires forming a single backward pass over a logical batch of size $M_i = 2048$, which is well above the typical mini-batch size used during training. On clients whose accelerators cannot hold $M_i$ samples in memory, the masking step is implemented with gradient accumulation: $D'_i$ is processed as $M_i / b$ micro-batches of size $b$, each contributes $\nabla_{\theta} \mathcal{L}\bigl(G_{\mathrm{adv}}(B'_{j}; \theta_i)\bigr)$, and the accumulated sum is applied as a single masking step before the next epoch begins. Because the mean-loss gradient is linear in the per-sample gradients, gradient accumulation yields the same $g'^{(e)}_i$ as a single large-batch pass, and the linear-leakage collision argument is unchanged: each accumulated micro-batch deposits its $M_i / (M_i/b) = b$ contributions into the same leakage bins as the synchronous formulation. In practice this only changes wall-clock time, not the privacy and utility behavior of the defense.

\textbf{Generator requirements (not architecture-agnostic).} The procedure does not depend on a specific generative-model family, but it does depend on the auxiliary samples being sufficiently task-relevant: samples that are far out-of-distribution either fail to enter the same leakage bins as $D_i$ (so they do not induce collisions) or pull the model away from the real data distribution (hurting utility). In our experiments we instantiate $\mathcal{G}$ with a publicly available latent diffusion model conditioned on locally known classes; conditional GANs, conditional VAEs, or domain-specific clinical generators are also viable as long as the in-distribution budget can be satisfied.

\textbf{In-distribution check.} The constraint $\mathbb{E}\lVert D_i - D'_i\rVert^2 \leq H$ is implemented per-client using a small validation split of $D_i$ and a simple distance estimator (e.g., feature-space MSE in a frozen feature extractor); samples whose addition would exceed $H$ are rejected. In practice this filter triggers infrequently when $\mathcal{G}$ is reasonably well matched to the modality.

\textbf{Compatibility with secure aggregation.} \sysname operates entirely client-side and does not modify the FL communication protocol. When SA is enabled, $\delta_i^{t}$ is fed through the SA mask exactly as a standard update, and all guarantees of the underlying SA protocol continue to hold; the privacy benefit from \sysname is additive.

\section{Additional Experimental Details}
\label{appendix:exp_details}

This section records the implementation choices abbreviated in Sec.~\ref{sec:exp-setup}. The main text keeps only the experimental essentials; the details below specify the exact simulation, models, generation pipeline, baselines, and metrics.

\textbf{Evaluation scope and compute.} We evaluate the active-server linear-leakage threat model of~\cite{fowl2021robbing,zhao2023loki,shi2023scale}. Unless noted otherwise, quantitative results average multiple independent runs with different random seeds, including randomness from initialization and client participation. All timing and training experiments use a workstation with an Intel Core i9-11900K CPU and one NVIDIA GeForce RTX~3080 GPU; peak memory is governed by the logical masking batch $M_i$ when gradient accumulation is used.

\textbf{Federated data and optimization.} We simulate $N=100$ clients with non-overlapping data partitions. To induce non-IID label skew, each client is restricted to examples from at most five classes, and the server samples $10\%$ of clients uniformly without replacement per FedAvg round. Local training uses Adam with learning rate $\eta=10^{-3}$, local epochs $E=3$, and mini-batch size $B=64$; gradient clipping is used only where required by a baseline implementation, such as DP calibration.

\textbf{Datasets and classifiers.} MNIST~\cite{lecun2010mnist} and CIFAR-10~\cite{krizhevsky2009learning} use a shared CNN with three $3{\times}3$ convolutional layers (32 filters, ReLU, $2{\times}2$ max-pooling), a 512-unit fully connected layer, and a 10-class output layer. For ChestMNIST, OrganAMNIST, and PathMNIST~\cite{yang2023medmnist}, we use the same backbone, adapt the input channel count to the dataset, and set the output width to the dataset-native label cardinality.

\textbf{Defense data and hyperparameters.} Defense images are generated with Stable Diffusion~2.0~\cite{Rombach_2022_CVPR} using OpenCLIP ViT-H/14. Clients use class-aligned prompts derived from locally visible labels, resize or crop generated candidates to match the dataset tensor layout, and filter them by the in-distribution budget $\mathbb{E}\lVert D_i-D'_i\rVert^2\leq H$ estimated on a small held-out slice of $D_i$. Unless swept, \sysname{} uses $M_i=2048$; Soteria and GC use pruning ratio $0.8$; DP uses Gaussian noise with $\varepsilon=1.0$ and sensitivity $\Delta=10^{-4}$; GD and Outpost follow the hyperparameters released with their reference implementations~\cite{wang2022protect,wang2023more}. One accepted defense candidate takes roughly five seconds to generate on our workstation, including decoding and filtering.

\textbf{Attack protocol and metrics.} Robbing-the-Fed, LOKI, and Scale-MIA use the same FL schedule $(E,B)$ as the utility experiments. Reconstructions are evaluated from the aggregated global model after the first FL round unless a figure explicitly sweeps another variable; crafted FC leakage layers use the attack papers' recommended width, $k=1024$, unless otherwise stated. We report PSNR, SSIM, and reconstruction rate (RR), where RR counts images with $\mathrm{PSNR}>18~\mathrm{dB}$ as successful reconstructions~\cite{shi2023scale,fowl2021robbing}. Parenthetical PSNR/SSIM values in tables average only over successful reconstructions.

\section{Robbing-the-Fed under Varying Local Training Hyperparameters}
\label{appendix:local_robbing}

Tab.~\ref{tab:local} supplements Sec.~\ref{sec:threat-model} by varying local batch size, learning rate, and number of local epochs under Robbing-the-Fed~\cite{fowl2021robbing}. The tested utility-preserving hyperparameter changes do not materially reduce reconstruction quality, which motivates a defense mechanism that directly targets the leakage-capacity bottleneck.

\begin{table}[h!]
    \centering
    \caption{Sensitivity of Robbing-the-Fed~\cite{fowl2021robbing} to local training hyperparameters. The rows compare batch size, learning rate, and local-epoch choices that preserve normal training utility; attack quality remains high across these settings, showing that benign hyperparameter tuning alone is not a reliable defense.}
    \includegraphics[width=0.75\linewidth]{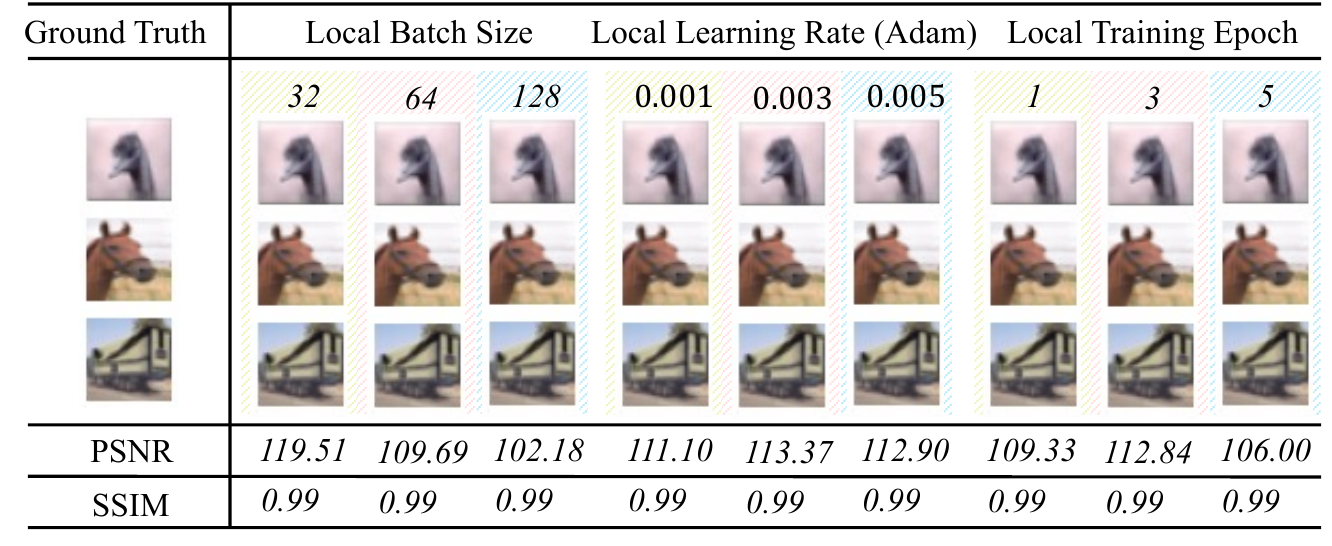}
    \label{tab:local}
    \vspace{-0.7cm}
\end{table}

\section{Additional Experiment: Multi-baseline Comparisons on CIFAR-10}
\label{appendix:cifar_robbing_additional}

This appendix complements the medical-imaging results in Sec.~\ref{EE} with CIFAR-10 reconstructions against a broader defense set. We compare \sysname with DP~\cite{46432}, Outpost~\cite{wang2023more}, GD~\cite{wang2022protect}, Soteria~\cite{sun2021soteria}, and GC~\cite{zhu2019deep}, using Robbing-the-Fed~\cite{fowl2021robbing} as the representative high-fidelity linear-leakage attack.

\begin{table*}[h!]
  \centering
  \caption{CIFAR-10 reconstructions by Robbing-the-Fed under six settings: no defense, \sysname, DP, GC, Soteria, and Outpost. RR is the fraction of samples with PSNR$>$18. Blue parenthetical PSNR/SSIM values average only over successfully reconstructed samples, separating overall protection from the fidelity of the few surviving reconstructions.}
  \label{tab:recover-cifar-appendix}

  \renewcommand{\arraystretch}{1.15}
  \setlength{\tabcolsep}{10pt}

  \begin{tabular}{cccc}
  \hline
  \textbf{Ground Truth} &
  \textbf{No defense} &
  \textbf{\sysname} &
  \textbf{DP} \\
  \hline

  \includegraphics[width=0.20\linewidth]{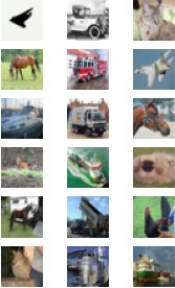} &
  \includegraphics[width=0.20\linewidth]{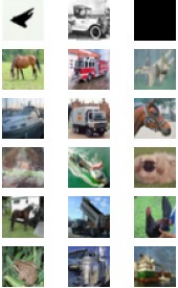} &
  \includegraphics[width=0.20\linewidth]{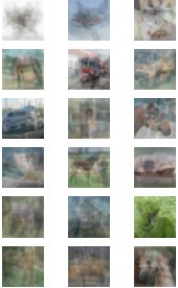} &
  \includegraphics[width=0.20\linewidth]{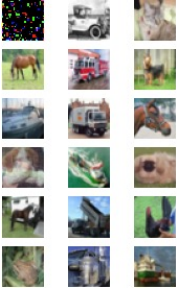} \\
  \hline
  \vspace{1mm}

  \textbf{RR (\%)} &
  87.50 &
  12.50 &
  84.38 \\

  \hline

  \textbf{PSNR} &
  99.24 {\scriptsize\textcolor{blue}{(112.02)}} &
  16.30 {\scriptsize\textcolor{blue}{(28.30)}} &
  58.99 {\scriptsize\textcolor{blue}{(63.01)}} \\

  \hline

  \textbf{SSIM} &
  0.96 {\scriptsize\textcolor{blue}{(0.99)}} &
  0.40 {\scriptsize\textcolor{blue}{(0.99)}} &
  0.96 {\scriptsize\textcolor{blue}{(0.99)}} \\

  \hline
  \hline

  \textbf{Ground Truth} &
  \textbf{GC} &
  \textbf{Soteria} &
  \textbf{Outpost} \\
  \hline

  \includegraphics[width=0.20\linewidth]{figures/18gt.pdf} &
  \includegraphics[width=0.20\linewidth]{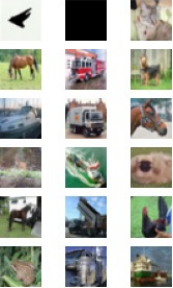} &
  \includegraphics[width=0.20\linewidth]{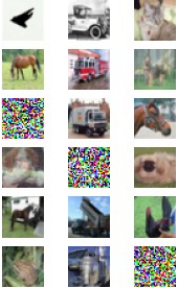} &
  \includegraphics[width=0.20\linewidth]{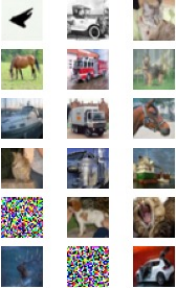} \\
  \hline
  \vspace{1mm}

  \textbf{RR (\%)} &
  64.06 &
  81.25 &
  79.69 \\

  \hline

  \textbf{PSNR} &
  75.03 {\scriptsize\textcolor{blue}{(112.20)}} &
  53.57 {\scriptsize\textcolor{blue}{(57.14)}} &
  54.44 {\scriptsize\textcolor{blue}{(57.58)}} \\

  \hline

  \textbf{SSIM} &
  0.95 {\scriptsize\textcolor{blue}{(0.99)}} &
  0.95 {\scriptsize\textcolor{blue}{(0.99)}} &
  0.95 {\scriptsize\textcolor{blue}{(0.99)}} \\

  \hline
  \end{tabular}
\end{table*}

Tab.~\ref{tab:recover-cifar-appendix} reports the comparison. \sysname reduces RR to $12.50\%$ from $87.50\%$ without defense, and the remaining reconstructions are visually blended. DP, GC, Soteria, and Outpost reduce some aggregate metrics, but their successful reconstructions retain high conditional PSNR/SSIM, indicating that recognizable inputs can still survive. These CIFAR-10 results are consistent with the medical-imaging results in Sec.~\ref{EE}: increasing the effective batch size through synthetic masking directly disrupts the closed-form leakage mechanism.

\FloatBarrier
\section{Additional Medical Utility and Privacy--Utility Results}
\label{appendix:medical}

\subsection{Medical Task Accuracy}
Table~\ref{tab:medical-accuracy} reports final medical test accuracy under the same 100-client non-IID FedAvg setting as Appendix~\ref{appendix:exp_details}. The Eloul et al.~\cite{eloul2024mixing} baseline uses MSE loss with same-label batching under the same FL and Robbing-the-Fed protocol. With $M=2048$, \sysname's accuracy differs from no defense by at most 0.3 percentage points across the three datasets.

\begin{table}[htbp]
    \centering
    \small
    \caption{Final medical test accuracy (\%). \sysname uses $M=2048$.}
    \label{tab:medical-accuracy}
    \renewcommand{\arraystretch}{1.1}
    \begin{tabular}{lrrr}
        \hline
        \textbf{Defense} & \textbf{ChestMNIST} & \textbf{OrganAMNIST} & \textbf{PathMNIST} \\
        \hline
        No defense & 55.8 & 58.6 & 57.3 \\
        Eloul et al. & 55.7 & 58.5 & 57.2 \\
        \sysname & 55.5 & 58.8 & 57.2 \\
        Soteria & 53.8 & 56.6 & 55.3 \\
        GC & 51.9 & 54.7 & 53.4 \\
        GD & 47.2 & 50.0 & 48.7 \\
        Outpost & 49.0 & 51.8 & 50.5 \\
        DP & 44.7 & 47.5 & 46.2 \\
        \hline
    \end{tabular}
\end{table}

\FloatBarrier
\subsection{Privacy--Utility Trade-off}
Table~\ref{tab:medical-tradeoff} reports the masking-batch sweep and the Eloul et al. baseline against Robbing-the-Fed, with $B=64$, $E=3$, and $k=1024$. Increasing $M$ reduces reconstruction rate, while accuracy stays within 0.4 percentage points of no defense.

\begin{table}[htbp]
    \centering
    \small
    \caption{Medical privacy--utility trade-off against Robbing-the-Fed. PSNR (dB) and SSIM are overall reconstruction metrics.}
    \label{tab:medical-tradeoff}
    \renewcommand{\arraystretch}{1.1}
    \begin{tabular}{lrrrr}
        \hline
        \textbf{Setting} & \textbf{ACC (\%)} $\uparrow$ & \textbf{RR (\%)} $\downarrow$ & \textbf{PSNR} $\downarrow$ & \textbf{SSIM} $\downarrow$ \\
        \hline
        \multicolumn{5}{l}{\textbf{ChestMNIST}} \\
        No defense & 55.8 & 89.06 & 62.40 & 0.94 \\
        Eloul et al. & 55.7 & 85.94 & 58.90 & 0.92 \\
        \sysname, $M=512$ & 56.0 & 49.95 & 34.59 & 0.69 \\
        \sysname, $M=1024$ & 55.9 & 27.03 & 25.06 & 0.55 \\
        \sysname, $M=1536$ & 55.7 & 16.24 & 21.08 & 0.49 \\
        \sysname, $M=2048$ & 55.5 & 9.50 & 18.70 & 0.46 \\
        \hline
        \multicolumn{5}{l}{\textbf{OrganAMNIST}} \\
        No defense & 58.6 & 82.81 & 50.60 & 0.89 \\
        Eloul et al. & 58.5 & 79.69 & 47.90 & 0.87 \\
        \sysname, $M=512$ & 58.8 & 47.82 & 31.19 & 0.69 \\
        \sysname, $M=1024$ & 59.0 & 27.31 & 24.54 & 0.58 \\
        \sysname, $M=1536$ & 58.9 & 17.66 & 21.76 & 0.53 \\
        \sysname, $M=2048$ & 58.8 & 11.63 & 20.10 & 0.51 \\
        \hline
        \multicolumn{5}{l}{\textbf{PathMNIST}} \\
        No defense & 57.3 & 78.13 & 45.20 & 0.86 \\
        Eloul et al. & 57.2 & 75.00 & 42.60 & 0.83 \\
        \sysname, $M=512$ & 57.5 & 44.34 & 27.45 & 0.61 \\
        \sysname, $M=1024$ & 57.6 & 24.53 & 21.36 & 0.47 \\
        \sysname, $M=1536$ & 57.4 & 15.21 & 18.82 & 0.41 \\
        \sysname, $M=2048$ & 57.2 & 9.38 & 17.30 & 0.38 \\
        \hline
    \end{tabular}
\end{table}

\FloatBarrier
\section{Additional Sensitivity Experiments}
\label{appendix:sensitivity}

\subsection{Generator Ablation}
\label{appendix:generator}
Table~\ref{tab:generator} compares Stable Diffusion 2.0 with SDXL 1.0~\cite{podell2023sdxl}, VQ-VAE-2~\cite{razavi2019vqvae2}, and StyleGAN2-ADA~\cite{karras2020ada}, using the same $M=2048$ cache and FL/Robbing-the-Fed configuration ($B=64$, $E=3$, $k=1024$). The default public Stable Diffusion checkpoint is used without dataset-specific fine-tuning; alternatives are conditioned or adapted on a sanctioned client-side training split where needed. All variants use the same locally visible labels, preprocessing, and acceptance filter, with no test images used for generation or adaptation. Accuracy and reconstruction metrics remain close across the evaluated generators.

\begin{table}[htbp]
    \centering
    \small
    \caption{Generator ablation under the same \sysname and Robbing-the-Fed configuration.}
    \label{tab:generator}
    \renewcommand{\arraystretch}{1.1}
    \begin{tabular}{lrrrr}
        \hline
        \textbf{Setting} & \textbf{ACC (\%)} $\uparrow$ & \textbf{RR (\%)} $\downarrow$ & \textbf{PSNR} $\downarrow$ & \textbf{SSIM} $\downarrow$ \\
        \hline
        \multicolumn{5}{l}{\textbf{ChestMNIST}} \\
        No defense & 55.8 & 89.06 & 62.40 & 0.94 \\
        Stable Diffusion 2.0 & 55.5 & 9.50 & 18.70 & 0.46 \\
        SDXL 1.0 & 55.6 & 9.06 & 18.30 & 0.44 \\
        VQ-VAE-2 & 55.4 & 10.63 & 19.20 & 0.48 \\
        StyleGAN2-ADA & 55.3 & 11.25 & 19.60 & 0.49 \\
        \hline
        \multicolumn{5}{l}{\textbf{OrganAMNIST}} \\
        No defense & 58.6 & 82.81 & 50.60 & 0.89 \\
        Stable Diffusion 2.0 & 58.8 & 11.63 & 20.10 & 0.51 \\
        SDXL 1.0 & 58.7 & 10.94 & 19.60 & 0.49 \\
        VQ-VAE-2 & 58.6 & 12.50 & 20.70 & 0.53 \\
        StyleGAN2-ADA & 58.5 & 12.81 & 21.10 & 0.54 \\
        \hline
        \multicolumn{5}{l}{\textbf{PathMNIST}} \\
        No defense & 57.3 & 78.13 & 45.20 & 0.86 \\
        Stable Diffusion 2.0 & 57.2 & 9.38 & 17.30 & 0.38 \\
        SDXL 1.0 & 57.3 & 8.75 & 16.90 & 0.36 \\
        VQ-VAE-2 & 57.1 & 10.31 & 17.90 & 0.41 \\
        StyleGAN2-ADA & 57.0 & 10.63 & 18.20 & 0.42 \\
        \hline
    \end{tabular}
\end{table}

\FloatBarrier
\subsection{Non-IID Partitions}
\label{appendix:noniid}
Following Li et al.~\cite{li2022noniid}, Table~\ref{tab:noniid} reports distribution-based label skew, two-class label skew, feature skew, and quantity skew. All settings retain $B=64$, $M=2048$, $k=1024$, and 10\% client participation. The two-class constraint yields the lowest accuracy, while reconstruction rates remain close across partitions.

\begin{table}[htbp]
    \centering
    \small
    \caption{Non-IID sensitivity of \sysname against Robbing-the-Fed. Partition labels follow the experimental settings.}
    \label{tab:noniid}
    \renewcommand{\arraystretch}{1.1}
    \begin{tabular}{lrrrr}
        \hline
        \textbf{Setting} & \textbf{ACC (\%)} $\uparrow$ & \textbf{RR (\%)} $\downarrow$ & \textbf{PSNR} $\downarrow$ & \textbf{SSIM} $\downarrow$ \\
        \hline
        \multicolumn{5}{l}{\textbf{ChestMNIST}} \\
        Original: at most 5 classes & 55.5 & 9.50 & 18.70 & 0.46 \\
        Label Dir(0.5) & 58.8 & 9.38 & 18.61 & 0.45 \\
        Label Dir(0.1) & 53.8 & 9.62 & 18.82 & 0.47 \\
        Label quantity: 2 classes & 43.6 & 9.71 & 18.91 & 0.47 \\
        Feature Gau(0.1) & 59.5 & 9.94 & 19.05 & 0.48 \\
        Quantity Dir(0.5) & 60.2 & 9.44 & 18.66 & 0.46 \\
        \hline
        \multicolumn{5}{l}{\textbf{OrganAMNIST}} \\
        Original: at most 5 classes & 58.8 & 11.63 & 20.10 & 0.51 \\
        Label Dir(0.5) & 62.1 & 11.48 & 19.95 & 0.50 \\
        Label Dir(0.1) & 57.0 & 11.78 & 20.25 & 0.52 \\
        Label quantity: 2 classes & 46.4 & 11.91 & 20.37 & 0.52 \\
        Feature Gau(0.1) & 62.8 & 12.08 & 20.51 & 0.53 \\
        Quantity Dir(0.5) & 63.5 & 11.57 & 20.04 & 0.51 \\
        \hline
        \multicolumn{5}{l}{\textbf{PathMNIST}} \\
        Original: at most 5 classes & 57.2 & 9.38 & 17.30 & 0.38 \\
        Label Dir(0.5) & 60.5 & 9.25 & 17.18 & 0.37 \\
        Label Dir(0.1) & 55.4 & 9.51 & 17.42 & 0.39 \\
        Label quantity: 2 classes & 44.8 & 9.62 & 17.54 & 0.39 \\
        Feature Gau(0.1) & 61.2 & 9.78 & 17.72 & 0.40 \\
        Quantity Dir(0.5) & 61.9 & 9.34 & 17.28 & 0.38 \\
        \hline
    \end{tabular}
\end{table}

\FloatBarrier
\subsection{Classifier Backbones}
\label{appendix:backbones}
Table~\ref{tab:backbone} compares classifier backbones under the original non-IID partition, with $M=2048$ and the same crafted FC leakage front-end. All classifiers use dataset-native resolution without classifier pretraining; ViT-S/4 uses $4\times4$ patches. Backbone changes affect accuracy, while the reported reconstruction metrics remain close.

\begin{table}[htbp]
    \centering
    \small
    \caption{Backbone sensitivity under the same \sysname and Robbing-the-Fed setting.}
    \label{tab:backbone}
    \renewcommand{\arraystretch}{1.1}
    \begin{tabular}{lrrrr}
        \hline
        \textbf{Setting} & \textbf{ACC (\%)} $\uparrow$ & \textbf{RR (\%)} $\downarrow$ & \textbf{PSNR} $\downarrow$ & \textbf{SSIM} $\downarrow$ \\
        \hline
        \multicolumn{5}{l}{\textbf{ChestMNIST}} \\
        3-layer CNN & 55.5 & 9.50 & 18.70 & 0.46 \\
        ResNet-18 & 57.8 & 9.47 & 18.68 & 0.46 \\
        ViT-S/4 & 56.4 & 9.54 & 18.74 & 0.46 \\
        Swin-T & 57.2 & 9.49 & 18.69 & 0.46 \\
        \hline
        \multicolumn{5}{l}{\textbf{OrganAMNIST}} \\
        3-layer CNN & 58.8 & 11.63 & 20.10 & 0.51 \\
        ResNet-18 & 61.4 & 11.59 & 20.07 & 0.51 \\
        ViT-S/4 & 59.9 & 11.67 & 20.14 & 0.51 \\
        Swin-T & 60.7 & 11.61 & 20.09 & 0.51 \\
        \hline
        \multicolumn{5}{l}{\textbf{PathMNIST}} \\
        3-layer CNN & 57.2 & 9.38 & 17.30 & 0.38 \\
        ResNet-18 & 60.1 & 9.35 & 17.27 & 0.38 \\
        ViT-S/4 & 58.4 & 9.42 & 17.35 & 0.38 \\
        Swin-T & 59.4 & 9.37 & 17.29 & 0.38 \\
        \hline
    \end{tabular}
\end{table}

\FloatBarrier

\section{Convergence Analysis: Assumptions and Theorem}
\label{appendix:convergence}

This section contains the full assumptions, intermediate variance bounds, and convergence theorem referenced from the system design. The key convergence issue is the discrepancy between private data $D_i$ and synthetic defense data $D'_i$: under standard convex federated-learning assumptions~\cite{Li2020On}, the budget $\mathbb{E}\lVert D_i - D'_i\rVert^2 \leq H$ enters the variance and norm constants but does not change the asymptotic $\mathcal{O}(1/T)$ rate. Our analysis extends the FedAvg framework of~\cite{Li2020On} to the \sysname setting; the only structural addition is the data-discrepancy bound between the private dataset $D_i$ and the synthetic defense dataset $D'_i$, formalized in Assumption~\ref{as:6}.

\textbf{Scope.} \emph{This analysis is not intended as a full convergence guarantee for the non-convex neural networks used in our experiments.} It is a stability argument under the same standard convex assumptions used in~\cite{Li2020On} (smoothness, strong convexity, bounded stochastic gradients, uniform client sampling). Its purpose is to make explicit how the auxiliary-gradient discrepancy $H = \mathbb{E}\lVert D_i - D'_i\rVert^2$ propagates into the variance and uniform-norm constants of the bound. Strong convexity is well known not to hold for deep neural networks, so the theorem below should be read as informing the design choice (smaller $H$ tightens constants) rather than as a deep-learning convergence result. Empirical utility on deep models is reported in Sec.~\ref{sec:experiments}.

\subsection*{Assumptions}

\begin{assumption} \label{as:1}
\textit{Local objective functions} $F_1, \dots, F_N$ \textit{are all $L$-smooth: for all $v$ and $w$,} $F_i(v) \leq F_i(w) + (v-w)^T\nabla F_i(w) + \frac{L}{2}\left\lVert v-w \right\rVert^2_2$.
\end{assumption}

\begin{assumption} \label{as:2}
$F_1, \dots, F_N$ \textit{are all $\mu$-strongly convex: for all $v$ and $w$,} $F_i(v) \geq F_i(w) + (v-w)^T\nabla F_i(w) + \frac{\mu}{2}\left\lVert v-w \right\rVert^2_2$.
\end{assumption}

\begin{assumption} \label{as:3}
\textit{Let $D_i$ be sampled uniformly at random from the $i$-th device's local data. The variance of stochastic gradients on each device is bounded:} $\mathbb{E}\lVert \nabla F_i(D_i;\theta_i^t) - \nabla F_i(\theta_i^t)\rVert^2 \leq \sigma^2_i$ \textit{for $i=1,\dots,N$.}
\end{assumption}

\begin{assumption} \label{as:4}
\textit{The expected squared norm of stochastic gradients is uniformly bounded: $\mathbb{E}\lVert \nabla F_i(D_i;\theta_i^t)\rVert^2 \leq G^2$ for all $i=1,\dots,N$ and $t=1,\dots,T-1$.}
\end{assumption}

\begin{assumption} \label{as:5}
\textit{Let $S_t$ contain $K$ indices sampled uniformly without replacement from $[N]$. Assume balanced data: $p_1 = \dots = p_N = 1/N$. The aggregation step of FedAvg is $\theta_t \leftarrow \frac{N}{K}\sum_{i \in S_t} p_i \theta_i^t$.}
\end{assumption}

\begin{assumption} \label{as:6}
\textit{(Defense-data discrepancy.) The expected squared distance between the private data $D_i$ and the synthesized defense data $D'_i$ is bounded:} $\mathbb{E}\lVert D_i - D'_i \rVert^2 \leq H$ \textit{for all $i=1,\dots,N$ and $t=1,\dots,T-1$.}
\end{assumption}

\subsection*{Gradient variance under mixed private and defense data}

We derive the new variance bound replacing Assumption~\ref{as:3}. Write the per-round effective update direction as $\delta^t_i = (1-\lambda)\nabla F_i(D_i;\theta_i^t) + \lambda \nabla F_i(D'_i;\theta_i^t)$ with $\lambda \in [0,1]$ controlled by $M_i / (B + M_i)$. Using Assumptions~\ref{as:1} and~\ref{as:6},
\begin{align}
    & \mathbb{E}\left\lVert \delta^t_i - \nabla F_i(\theta_i^t)\right\rVert^2 \notag \\
    &\leq \mathbb{E}\left\lVert \nabla F_i(D_i;\theta_i^t) - \nabla F_i(\theta_i^t)\right\rVert^2 + \mathbb{E}\left\lVert \delta^t_i - \nabla F_i(D_i;\theta_i^t)\right\rVert^2 \notag \\
    &\leq \sigma^2_i + \mathbb{E} \left\lVert \delta^t_i - \nabla F_i(D_i;\theta_i^t) \right\rVert^2 \notag \\
    &= \sigma^2_i +\mathbb{E} \left\lVert \lambda (\nabla F_i(D'_i;\theta_i^t) - \nabla F_i(D_i;\theta_i^t)) \right\rVert^2 \notag \\
    &\leq \sigma^2_i + \mathbb{E} \left\lVert \nabla F_i(D'_i;\theta_i^t) - \nabla F_i(D_i;\theta_i^t) \right\rVert^2 \notag \\
    &\leq \sigma^2_i + L\,\mathbb{E}\left\lVert D_i - D'_i \right\rVert^2 \notag \\
    &\leq \sigma^2_i + LH.
\end{align}
By the same derivation, the uniform bound in Assumption~\ref{as:4} becomes $\mathbb{E}\lVert \nabla F_i(D_i;\theta_i^t)\rVert^2 \leq G^2 + LH$. Define $F^*$ and $F^*_i$ as the minimum of $F$ and $F_i$, and set $\Gamma = F^* - \sum^N_{i=1} p_i F^*_i$.

\subsection*{Convergence theorem}

\begin{theorem}
\textit{Under Assumptions~\ref{as:1}--\ref{as:6}, with $\kappa = L/\mu$, $\gamma = \max\{8\kappa, E\}$, and the diminishing learning-rate schedule $\eta_t = 2/(\mu(\gamma + t))$, after $T$ communication rounds}
\begin{equation}
    \mathbb{E}[F(\theta^{T})] - F^* \;\leq\; \frac{\kappa}{\gamma + T - 1}\left(\frac{2(B+C)}{\mu} + \frac{\mu\gamma}{2}\,\mathbb{E}\left\lVert \theta^{1} - \theta^* \right\rVert^2\right),
\end{equation}
\textit{where}
\begin{align}
    B &= \sum^N_{i=1} p^2_i (\sigma^2_i + LH) + 6L\Gamma + 8(E-1)^2 (G^2 + LH), \\
    C &= \frac{N-K}{N-1}\,\frac{4}{K}\,E^2 (G^2 + LH).
\end{align}
\end{theorem}

\textit{Remark.} Setting $H = 0$ (i.e., $D'_i$ identical in distribution to $D_i$) recovers the bound of~\cite{Li2020On}, so \sysname inherits the standard $\mathcal{O}(1/T)$ rate; $H > 0$ enlarges the constants $B$ and $C$ by a controlled amount but does not affect the rate.

\section{Survey of Gradient-based MIAs in Federated Learning}
\label{appendix:mia_survey}

Tab.~\ref{tab:literature-review} summarizes representative MIAs by whether they break SA, the batch-size constraint at which they remain effective, and whether they rely on iterative optimization. The table substantiates the introduction's central observation: the local batch size relative to the first-FC-layer capacity is the unifying bottleneck across the entire literature, and \sysname turns this bottleneck into a defense.

\begin{table}[h!]
    \centering
    \small
    \setlength{\tabcolsep}{4pt}
    \caption{Representative gradient-based MIAs in FL. ``Break SA'' indicates attacks demonstrated under secure aggregation or equivalent aggregate-only settings. ``Batch Size Constraint'' reports the largest batch scale at which the attack is typically effective under a first crafted FC layer with 1024 neurons, highlighting why leakage capacity is the relevant bottleneck for \sysname.}
    \label{tab:literature-review}
    \begin{tabular}{lccc}
        \hline
        \textbf{MIA} & \textbf{Break SA} & \textbf{Batch Size} & \textbf{Optimization} \\
        \hline
        DLG \cite{zhu2019deep} & No & 1 & Yes \\
        iDLG \cite{zhao2020idlg} & No & 1 & Yes \\
        Inverting Gradients~\cite{geiping2020inverting} & No & 8 & Yes \\
        GradInversion~\cite{yin2021see} & No & 48 & Yes \\
        GradViT~\cite{hatamizadeh2022gradvit} & No & 8 & Yes \\
        APRIL-Optim~\cite{lu2022april} & No & 1 & Yes \\
        APRIL-Analytic~\cite{lu2022april} & No & 1 & Yes \\
        R-GAP~\cite{zhu2020r} & No & 1 & Yes \\
        Fishing for Data~\cite{wen2022fishing} & Yes & 256 & Yes \\
        Eluding SA~\cite{pasquini2022eluding} & Yes & 512 & Yes \\
        Robbing-the-Fed~\cite{fowl2021robbing} & Yes & 1024 & No \\
        LOKI~\cite{zhao2023loki} & Yes & 1024 & No \\
        Scale-MIA~\cite{shi2023scale} & Yes & 1024 & No \\
        \hline
    \end{tabular}
\end{table}

\section{Extended Discussion and Limitations}
\label{appendix:discussion}
\label{discussion}

\textbf{Deployment assumptions.} \sysname is designed for the server-side MIA threat model studied in recent linear-leakage attacks~\cite{fowl2021robbing,zhao2023loki,shi2023scale}: clients follow the FL protocol, while the server may broadcast a model crafted for reconstruction. Client-side poisoning, backdoors, and malicious client behavior are orthogonal FL security problems and can be combined with \sysname through existing robust-aggregation or poisoning-defense tools.

\textbf{Generator choice.} \sysname only requires a class- or text-conditional generator whose samples are close enough to the client modality to enter the same leakage bins as real data. Public pretrained generators, domain-adapted generators, or generators fine-tuned on sanctioned in-domain corpora are all compatible. The defense data remains local and is used only for masking-gradient computation, so it is not revealed to the server. Better-matched generators reduce the in-distribution budget $H$ in Appendix~\ref{appendix:convergence}, which helps preserve utility while maintaining bin collisions.

\textbf{Hyperparameters and resources.} The main defense parameters are $M_i$ and $H$. $M_i$ should make $B+M_i$ exceed the leakage-bin capacity $k$, so the required defense batch scales with the capacity of the crafted layer; our default $M_i=2048$ is larger than the $k=1024$ setting used by the evaluated linear-leakage attacks. $H$ is selected with a small client-side validation split to balance synthetic diversity and utility. When $M_i$ does not fit in memory, clients can compute the same logical masking gradient by accumulating micro-batches; this preserves the collision argument and trades memory for wall-clock time.

\textbf{Adaptive servers.} The evaluated attacks cover the known scalable linear-leakage family. A server could increase the crafted FC width, in which case \sysname requires $M_i$ to scale with the larger leakage capacity; choosing $M_i$ under unknown or adaptive $k$ is therefore a deployment consideration. A server could also try to estimate and remove synthetic contributions, but each leakage bin mixes unknown real and synthetic samples, making demixing underdetermined without client-side sample identities. These observations do not preclude future attack variants, but they show that \sysname targets the structural bottleneck shared by current closed-form attacks.

\textbf{Scope of the evidence.} The convergence result is a stability statement under standard convex assumptions for the federated setting; empirical utility in the non-convex deep-learning regime is evaluated directly in Sec.~\ref{sec:experiments}. The medical experiments cover three MedMNIST modalities at benchmark resolution. Applying the same mechanism to higher-resolution volumes or whole-slide pathology mainly changes the generator and compute requirements, not the client-side masking protocol.

\newpage
\section*{NeurIPS Paper Checklist}

\begin{enumerate}

\item {\bf Claims}
    \item[] Question: Do the main claims made in the abstract and introduction accurately reflect the paper's contributions and scope?
    \item[] Answer: \answerYes{}
    \item[] Justification: The abstract and Sec.~\ref{introduction} state the MIA threat model, the proposed \sysname{} defense, and the standard/medical benchmark evaluation. The empirical claims are supported in Sec.~\ref{EE}, and the theory claim is scoped in Appendix~\ref{appendix:convergence}.
    \item[] Guidelines:
    \begin{itemize}
        \item The answer \answerNA{} means that the abstract and introduction do not include the claims made in the paper.
        \item The abstract and/or introduction should clearly state the claims made, including the contributions made in the paper and important assumptions and limitations. A \answerNo{} or \answerNA{} answer to this question will not be perceived well by the reviewers.
        \item The claims made should match theoretical and experimental results, and reflect how much the results can be expected to generalize to other settings.
        \item It is fine to include aspirational goals as motivation as long as it is clear that these goals are not attained by the paper.
    \end{itemize}

\item {\bf Limitations}
    \item[] Question: Does the paper discuss the limitations of the work performed by the authors?
    \item[] Answer: \answerYes{}
    \item[] Justification: Appendix~\ref{appendix:discussion} discusses deployment assumptions, generator requirements, adaptive-server considerations, resource constraints, and the scope of the medical experiments. Sec.~\ref{sec:experiments} also notes that the reported medical studies use benchmark-resolution MedMNIST modalities.
    \item[] Guidelines:
    \begin{itemize}
        \item The answer \answerNA{} means that the paper has no limitation while the answer \answerNo{} means that the paper has limitations, but those are not discussed in the paper.
        \item The authors are encouraged to create a separate ``Limitations'' section in their paper.
        \item The paper should point out any strong assumptions and how robust the results are to violations of these assumptions (e.g., independence assumptions, noiseless settings, model well-specification, asymptotic approximations only holding locally). The authors should reflect on how these assumptions might be violated in practice and what the implications would be.
        \item The authors should reflect on the scope of the claims made, e.g., if the approach was only tested on a few datasets or with a few runs. In general, empirical results often depend on implicit assumptions, which should be articulated.
        \item The authors should reflect on the factors that influence the performance of the approach. For example, a facial recognition algorithm may perform poorly when image resolution is low or images are taken in low lighting. Or a speech-to-text system might not be used reliably to provide closed captions for online lectures because it fails to handle technical jargon.
        \item The authors should discuss the computational efficiency of the proposed algorithms and how they scale with dataset size.
        \item If applicable, the authors should discuss possible limitations of their approach to address problems of privacy and fairness.
        \item While the authors might fear that complete honesty about limitations might be used by reviewers as grounds for rejection, a worse outcome might be that reviewers discover limitations that aren't acknowledged in the paper. The authors should use their best judgment and recognize that individual actions in favor of transparency play an important role in developing norms that preserve the integrity of the community. Reviewers will be specifically instructed to not penalize honesty concerning limitations.
    \end{itemize}

\item {\bf Theory assumptions and proofs}
    \item[] Question: For each theoretical result, does the paper provide the full set of assumptions and a complete (and correct) proof?
    \item[] Answer: \answerYes{}
    \item[] Justification: Appendix~\ref{appendix:convergence} states Assumptions~\ref{as:1}--\ref{as:6}, derives the defense-specific variance terms, and gives the corresponding convergence theorem. We will provide any remaining standard proof details in the supplementary material while keeping the non-convex deep-learning scope explicit.
    \item[] Guidelines:
    \begin{itemize}
        \item The answer \answerNA{} means that the paper does not include theoretical results.
        \item All the theorems, formulas, and proofs in the paper should be numbered and cross-referenced.
        \item All assumptions should be clearly stated or referenced in the statement of any theorems.
        \item The proofs can either appear in the main paper or the supplemental material, but if they appear in the supplemental material, the authors are encouraged to provide a short proof sketch to provide intuition.
        \item Inversely, any informal proof provided in the core of the paper should be complemented by formal proofs provided in appendix or supplemental material.
        \item Theorems and Lemmas that the proof relies upon should be properly referenced.
    \end{itemize}

\item {\bf Experimental result reproducibility}
    \item[] Question: Does the paper fully disclose all the information needed to reproduce the main experimental results of the paper to the extent that it affects the main claims and/or conclusions of the paper (regardless of whether the code and data are provided or not)?
    \item[] Answer: \answerYes{}
    \item[] Justification: Sec.~\ref{sec:exp-setup} and Appendix~\ref{appendix:exp_details} specify the FL setup, datasets, model architecture, optimizer, baselines, defense batch size, attack protocol, and metrics. Sec.~\ref{EE} and Appendix~\ref{appendix:cifar_robbing_additional} report the main quantitative comparisons.
    \item[] Guidelines:
    \begin{itemize}
        \item The answer \answerNA{} means that the paper does not include experiments.
        \item If the paper includes experiments, a \answerNo{} answer to this question will not be perceived well by the reviewers: Making the paper reproducible is important, regardless of whether the code and data are provided or not.
        \item If the contribution is a dataset and\slash or model, the authors should describe the steps taken to make their results reproducible or verifiable.
        \item Depending on the contribution, reproducibility can be accomplished in various ways. For example, if the contribution is a novel architecture, describing the architecture fully might suffice, or if the contribution is a specific model and empirical evaluation, it may be necessary to either make it possible for others to replicate the model with the same dataset, or provide access to the model. In general. releasing code and data is often one good way to accomplish this, but reproducibility can also be provided via detailed instructions for how to replicate the results, access to a hosted model (e.g., in the case of a large language model), releasing of a model checkpoint, or other means that are appropriate to the research performed.
        \item While NeurIPS does not require releasing code, the conference does require all submissions to provide some reasonable avenue for reproducibility, which may depend on the nature of the contribution. For example
        \begin{enumerate}
            \item If the contribution is primarily a new algorithm, the paper should make it clear how to reproduce that algorithm.
            \item If the contribution is primarily a new model architecture, the paper should describe the architecture clearly and fully.
            \item If the contribution is a new model (e.g., a large language model), then there should either be a way to access this model for reproducing the results or a way to reproduce the model (e.g., with an open-source dataset or instructions for how to construct the dataset).
            \item We recognize that reproducibility may be tricky in some cases, in which case authors are welcome to describe the particular way they provide for reproducibility. In the case of closed-source models, it may be that access to the model is limited in some way (e.g., to registered users), but it should be possible for other researchers to have some path to reproducing or verifying the results.
        \end{enumerate}
    \end{itemize}

\item {\bf Open access to data and code}
    \item[] Question: Does the paper provide open access to the data and code, with sufficient instructions to faithfully reproduce the main experimental results, as described in supplemental material?
    \item[] Answer: \answerYes{}
    \item[] Justification: The datasets used in Sec.~\ref{sec:exp-setup} are public benchmarks and are cited, and the implementation-level procedure is documented in Algorithm~\ref{alg:aegis}, with further details in Appendix~\ref{appendix:implementation}. We will provide anonymized code and step-by-step reproduction instructions in the supplementary material.
    \item[] Guidelines:
    \begin{itemize}
        \item The answer \answerNA{} means that paper does not include experiments requiring code.
        \item Please see the NeurIPS code and data submission guidelines (\url{https://neurips.cc/public/guides/CodeSubmissionPolicy}) for more details.
        \item While we encourage the release of code and data, we understand that this might not be possible, so \answerNo{} is an acceptable answer. Papers cannot be rejected simply for not including code, unless this is central to the contribution (e.g., for a new open-source benchmark).
        \item The instructions should contain the exact command and environment needed to run to reproduce the results. See the NeurIPS code and data submission guidelines (\url{https://neurips.cc/public/guides/CodeSubmissionPolicy}) for more details.
        \item The authors should provide instructions on data access and preparation, including how to access the raw data, preprocessed data, intermediate data, and generated data, etc.
        \item The authors should provide scripts to reproduce all experimental results for the new proposed method and baselines. If only a subset of experiments are reproducible, they should state which ones are omitted from the script and why.
        \item At submission time, to preserve anonymity, the authors should release anonymized versions (if applicable).
        \item Providing as much information as possible in supplemental material (appended to the paper) is recommended, but including URLs to data and code is permitted.
    \end{itemize}

\item {\bf Experimental setting/details}
    \item[] Question: Does the paper specify all the training and test details (e.g., data splits, hyperparameters, how they were chosen, type of optimizer) necessary to understand the results?
    \item[] Answer: \answerYes{}
    \item[] Justification: Sec.~\ref{sec:exp-setup} gives the high-level experimental setup, and Appendix~\ref{appendix:exp_details} gives data partitioning, optimizer, model architecture, attack protocol, generation setup, and baseline hyperparameters. Metric definitions are provided in Sec.~\ref{sec:exp-setup} and Appendix~\ref{appendix:exp_details}.
    \item[] Guidelines:
    \begin{itemize}
        \item The answer \answerNA{} means that the paper does not include experiments.
        \item The experimental setting should be presented in the core of the paper to a level of detail that is necessary to appreciate the results and make sense of them.
        \item The full details can be provided either with the code, in appendix, or as supplemental material.
    \end{itemize}

\item {\bf Experiment statistical significance}
    \item[] Question: Does the paper report error bars suitably and correctly defined or other appropriate information about the statistical significance of the experiments?
    \item[] Answer: \answerYes{}
    \item[] Justification: Appendix~\ref{appendix:exp_details} states that quantitative metrics are averaged over multiple independent runs with different random seeds. We will report corresponding variability information, such as error bars or standard deviations, for the plotted and tabulated results.
    \item[] Guidelines:
    \begin{itemize}
        \item The answer \answerNA{} means that the paper does not include experiments.
        \item The authors should answer \answerYes{} if the results are accompanied by error bars, confidence intervals, or statistical significance tests, at least for the experiments that support the main claims of the paper.
        \item The factors of variability that the error bars are capturing should be clearly stated (for example, train/test split, initialization, random drawing of some parameter, or overall run with given experimental conditions).
        \item The method for calculating the error bars should be explained (closed form formula, call to a library function, bootstrap, etc.)
        \item The assumptions made should be given (e.g., Normally distributed errors).
        \item It should be clear whether the error bar is the standard deviation or the standard error of the mean.
        \item It is OK to report 1-sigma error bars, but one should state it. The authors should preferably report a 2-sigma error bar than state that they have a 96\% CI, if the hypothesis of Normality of errors is not verified.
        \item For asymmetric distributions, the authors should be careful not to show in tables or figures symmetric error bars that would yield results that are out of range (e.g., negative error rates).
        \item If error bars are reported in tables or plots, the authors should explain in the text how they were calculated and reference the corresponding figures or tables in the text.
    \end{itemize}

\item {\bf Experiments compute resources}
    \item[] Question: For each experiment, does the paper provide sufficient information on the computer resources (type of compute workers, memory, time of execution) needed to reproduce the experiments?
    \item[] Answer: \answerYes{}
    \item[] Justification: Appendix~\ref{appendix:exp_details} reports the CPU/GPU workstation used for timing and training experiments. Tab.~\ref{tab:running_time} reports wall-clock training time for representative MNIST and CIFAR-10 FL runs.
    \item[] Guidelines:
    \begin{itemize}
        \item The answer \answerNA{} means that the paper does not include experiments.
        \item The paper should indicate the type of compute workers CPU or GPU, internal cluster, or cloud provider, including relevant memory and storage.
        \item The paper should provide the amount of compute required for each of the individual experimental runs as well as estimate the total compute.
        \item The paper should disclose whether the full research project required more compute than the experiments reported in the paper (e.g., preliminary or failed experiments that didn't make it into the paper).
    \end{itemize}

\item {\bf Code of ethics}
    \item[] Question: Does the research conducted in the paper conform, in every respect, with the NeurIPS Code of Ethics \url{https://neurips.cc/public/EthicsGuidelines}?
    \item[] Answer: \answerYes{}
    \item[] Justification: The paper studies a privacy-preserving FL defense using public benchmark datasets, as described in Sec.~\ref{sec:exp-setup} and Appendix~\ref{appendix:exp_details}. The threat model and limitations in Sec.~\ref{sec:threat-model} and Appendix~\ref{appendix:discussion} address misuse risks from reconstruction attacks.
    \item[] Guidelines:
    \begin{itemize}
        \item The answer \answerNA{} means that the authors have not reviewed the NeurIPS Code of Ethics.
        \item If the authors answer \answerNo, they should explain the special circumstances that require a deviation from the Code of Ethics.
        \item The authors should make sure to preserve anonymity (e.g., if there is a special consideration due to laws or regulations in their jurisdiction).
    \end{itemize}

\item {\bf Broader impacts}
    \item[] Question: Does the paper discuss both potential positive societal impacts and negative societal impacts of the work performed?
    \item[] Answer: \answerYes{}
    \item[] Justification: Sec.~\ref{introduction} motivates the positive impact of protecting sensitive medical FL data. Sec.~\ref{sec:threat-model} and Appendix~\ref{appendix:discussion} discuss negative misuse risks from active servers and adaptive reconstruction attacks.
    \item[] Guidelines:
    \begin{itemize}
        \item The answer \answerNA{} means that there is no societal impact of the work performed.
        \item If the authors answer \answerNA{} or \answerNo, they should explain why their work has no societal impact or why the paper does not address societal impact.
        \item Examples of negative societal impacts include potential malicious or unintended uses (e.g., disinformation, generating fake profiles, surveillance), fairness considerations (e.g., deployment of technologies that could make decisions that unfairly impact specific groups), privacy considerations, and security considerations.
        \item The conference expects that many papers will be foundational research and not tied to particular applications, let alone deployments. However, if there is a direct path to any negative applications, the authors should point it out. For example, it is legitimate to point out that an improvement in the quality of generative models could be used to generate Deepfakes for disinformation. On the other hand, it is not needed to point out that a generic algorithm for optimizing neural networks could enable people to train models that generate Deepfakes faster.
        \item The authors should consider possible harms that could arise when the technology is being used as intended and functioning correctly, harms that could arise when the technology is being used as intended but gives incorrect results, and harms following from (intentional or unintentional) misuse of the technology.
        \item If there are negative societal impacts, the authors could also discuss possible mitigation strategies (e.g., gated release of models, providing defenses in addition to attacks, mechanisms for monitoring misuse, mechanisms to monitor how a system learns from feedback over time, improving the efficiency and accessibility of ML).
    \end{itemize}

\item {\bf Safeguards}
    \item[] Question: Does the paper describe safeguards that have been put in place for responsible release of data or models that have a high risk for misuse (e.g., pre-trained language models, image generators, or scraped datasets)?
    \item[] Answer: \answerNA{}
    \item[] Justification: The paper does not release a new high-capacity generative model, scraped dataset, or other high-risk asset. It uses public benchmarks and existing generative components, as described in Sec.~\ref{sec:exp-setup} and Appendix~\ref{appendix:exp_details}.
    \item[] Guidelines:
    \begin{itemize}
        \item The answer \answerNA{} means that the paper poses no such risks.
        \item Released models that have a high risk for misuse or dual-use should be released with necessary safeguards to allow for controlled use of the model, for example by requiring that users adhere to usage guidelines or restrictions to access the model or implementing safety filters.
        \item Datasets that have been scraped from the Internet could pose safety risks. The authors should describe how they avoided releasing unsafe images.
        \item We recognize that providing effective safeguards is challenging, and many papers do not require this, but we encourage authors to take this into account and make a best faith effort.
    \end{itemize}

\item {\bf Licenses for existing assets}
    \item[] Question: Are the creators or original owners of assets (e.g., code, data, models), used in the paper, properly credited and are the license and terms of use explicitly mentioned and properly respected?
    \item[] Answer: \answerYes{}
    \item[] Justification: Existing datasets, models, attacks, and defense baselines are credited through citations in \texttt{main.bib} and discussed in Sec.~\ref{sec:exp-setup} and Appendix~\ref{appendix:exp_details}. We will explicitly list the relevant licenses and terms of use for each asset in the supplementary material or final artifact documentation.
    \item[] Guidelines:
    \begin{itemize}
        \item The answer \answerNA{} means that the paper does not use existing assets.
        \item The authors should cite the original paper that produced the code package or dataset.
        \item The authors should state which version of the asset is used and, if possible, include a URL.
        \item The name of the license (e.g., CC-BY 4.0) should be included for each asset.
        \item For scraped data from a particular source (e.g., website), the copyright and terms of service of that source should be provided.
        \item If assets are released, the license, copyright information, and terms of use in the package should be provided. For popular datasets, \url{paperswithcode.com/datasets} has curated licenses for some datasets. Their licensing guide can help determine the license of a dataset.
        \item For existing datasets that are re-packaged, both the original license and the license of the derived asset (if it has changed) should be provided.
        \item If this information is not available online, the authors are encouraged to reach out to the asset's creators.
    \end{itemize}

\item {\bf New assets}
    \item[] Question: Are new assets introduced in the paper well documented and is the documentation provided alongside the assets?
    \item[] Answer: \answerNA{}
    \item[] Justification: The paper does not introduce a new public dataset, benchmark, or model asset. The contribution is a client-side defense mechanism described in Sec.~\ref{sec:system-design} and Algorithm~\ref{alg:aegis}.
    \item[] Guidelines:
    \begin{itemize}
        \item The answer \answerNA{} means that the paper does not release new assets.
        \item Researchers should communicate the details of the dataset\slash code\slash model as part of their submissions via structured templates. This includes details about training, license, limitations, etc.
        \item The paper should discuss whether and how consent was obtained from people whose asset is used.
        \item At submission time, remember to anonymize your assets (if applicable). You can either create an anonymized URL or include an anonymized zip file.
    \end{itemize}

\item {\bf Crowdsourcing and research with human subjects}
    \item[] Question: For crowdsourcing experiments and research with human subjects, does the paper include the full text of instructions given to participants and screenshots, if applicable, as well as details about compensation (if any)?
    \item[] Answer: \answerNA{}
    \item[] Justification: The experiments use public ML benchmarks and do not involve crowdsourcing or direct human-subjects studies. The datasets are listed in Sec.~\ref{sec:exp-setup} and Appendix~\ref{appendix:exp_details}.
    \item[] Guidelines:
    \begin{itemize}
        \item The answer \answerNA{} means that the paper does not involve crowdsourcing nor research with human subjects.
        \item Including this information in the supplemental material is fine, but if the main contribution of the paper involves human subjects, then as much detail as possible should be included in the main paper.
        \item According to the NeurIPS Code of Ethics, workers involved in data collection, curation, or other labor should be paid at least the minimum wage in the country of the data collector.
    \end{itemize}

\item {\bf Institutional review board (IRB) approvals or equivalent for research with human subjects}
    \item[] Question: Does the paper describe potential risks incurred by study participants, whether such risks were disclosed to the subjects, and whether Institutional Review Board (IRB) approvals (or an equivalent approval/review based on the requirements of your country or institution) were obtained?
    \item[] Answer: \answerNA{}
    \item[] Justification: No new human-subjects data were collected, and the experiments use public benchmark datasets listed in Sec.~\ref{sec:exp-setup}. Therefore IRB approval is not applicable to this study.
    \item[] Guidelines:
    \begin{itemize}
        \item The answer \answerNA{} means that the paper does not involve crowdsourcing nor research with human subjects.
        \item Depending on the country in which research is conducted, IRB approval (or equivalent) may be required for any human subjects research. If you obtained IRB approval, you should clearly state this in the paper.
        \item We recognize that the procedures for this may vary significantly between institutions and locations, and we expect authors to adhere to the NeurIPS Code of Ethics and the guidelines for their institution.
        \item For initial submissions, do not include any information that would break anonymity (if applicable), such as the institution conducting the review.
    \end{itemize}

\item {\bf Declaration of LLM usage}
    \item[] Question: Does the paper describe the usage of LLMs if it is an important, original, or non-standard component of the core methods in this research? Note that if the LLM is used only for writing, editing, or formatting purposes and does \emph{not} impact the core methodology, scientific rigor, or originality of the research, declaration is not required.
    \item[] Answer: \answerNA{}
    \item[] Justification: LLMs are not an original or non-standard component of the proposed defense or experiments. Defense-data synthesis uses conditional image generators, as described in Sec.~\ref{sec:system-design} and Sec.~\ref{sec:exp-setup}.

    \item[] Guidelines:
    \begin{itemize}
        \item The answer \answerNA{} means that the core method development in this research does not involve LLMs as any important, original, or non-standard components.
        \item Please refer to our LLM policy in the NeurIPS handbook for what should or should not be described.
    \end{itemize}

\end{enumerate}

\end{document}